\documentclass[manuscript]{aastex}

\usepackage{booktabs,fixltx2e}
\usepackage[flushleft]{threeparttable}

\shorttitle{New catalog of subdwarfs}
\shortauthors{Savcheva et al.}

\usepackage{rotating}

\begin{document}

\title{A New Sample of Cool Subdwarfs from SDSS: Properties and Kinematics}

\author{Antonia S. Savcheva\altaffilmark{1,2}, Andrew A. West\altaffilmark{2} and John J. Bochanski\altaffilmark{3}}
\affil{\altaffilmark{1}Harvard-Smithsonian Center for Astrophysics, 60 Garden Street, Cambridge, MA 02138, USA; \\
\altaffilmark{2}Astronomy Department, Boston University, 725 Commonwealth Ave., Boston, MA 02215, USA; \\ 
\altaffilmark{3}Department of Physics and Astronomy, Haverford College, 370 Lancaster Ave., Haverford, PA 19041}
\email{asavcheva@cfa.harvard.edu}

\begin{abstract}
We present a new sample of M subdwarfs compiled from the 7th data release of the Sloan Digital Sky Survey. With 3517 new subdwarfs, this new sample significantly increases the number of spectroscopically confirmed low-mass subdwarfs. This catalog also includes 905 extreme and 534 ultra sudwarfs. We present the entire catalog including observed and derived quantities, and template spectra created from co-added subdwarf spectra. We show color-color and reduced proper motion diagrams of the three metallicity classes, which are shown to separate from the disk dwarf population. The extreme and ultra subdwarfs are seen at larger values of reduced proper motion as expected for more dynamically heated populations. We determine 3D kinematics for all of the stars with proper motions. The color-magnitude diagrams show a clear separation of the three metallicity classes with the ultra and extreme subdwarfs being significantly closer to the main sequence than the ordinary subdwarfs. All subdwarfs lie below (fainter) and to the left (bluer) of the main sequence. Based on the average $(U,V,W)$ velocities and their dispersions, the extreme and ultra subdwarfs likely belong to the Galactic halo, while the ordinary subdwarfs are likely part of the old Galactic (or thick) disk. An extensive activity analysis of subdwarfs is performed using H$\alpha$ emission and 208 active subdwarfs are found. We show that while the activity fraction of subdwarfs rises with spectral class and levels off at the latest spectral classes, consistent with the behavior of M dwarfs, the extreme and ultra subdwarfs are basically flat.
 
\end{abstract}

\keywords{Stars: subdwarf --- Stars: statistics ---- Catalogs}

\section{Introduction}
M subdwarfs are low-mass ($0.08M_{\odot}<M<0.8M_{\odot}$), low-luminosity ($L<0.05L_{\odot}$) stars that are the metal-poor ([Fe/H]$\lesssim$-0.5) counterparts of cool, late-type dwarfs. Although M subdwarfs are not as abundant \citep[0.25\% of the Galactic stellar population;][]{reid05} as typical disk M dwarfs \citep[M dwarfs make up 70\% of the Galactic stellar population;][]{bochanski10}, they have similar properties, such as low temperatures and lifetimes greater than the Hubble time \citep{laughlin97}, making them excellent tracers of Galactic chemical and dynamical evolution. Thus, exploring populations of different metallicities helps probe the composition and evolution of the different components of the Galaxy, and the  Galactic merger history. Since some subdwarfs lie close to the hydrogen burning limit, they can be used to probe the lower end of the stellar mass function, extending it into the hydrogen-burning limit. In addition, the cool, dim atmospheres of these stars and their surroundings provide conditions for studying molecule and dust formation in low metallicity environments as well as radiative transfer in cool, metal-poor atmospheres, which cannot be tackled using metal-rich M dwarfs.  

M subdwarfs exhibit large metallicity-induced changes in their spectra relative to M dwarfs. The main difference between M dwarfs and subdwarfs is the strength of the TiO bands, which are much weaker in subdwarfs due to their low metallicities and hence low Ti and O abundances. As early as the 1970s metallic hydrides, such as MgH, FeH, and CaH have been used to identify subdwarfs  \citep{boeshaar76,mould76,mould78,bessell82}. The relative strengths of the TiO and CaH molecular bands have been traditionally used as a metallicity proxy for subdwarfs and spectral type indicator respectively. Different classification schemes have been devised to spectroscopically identify the metallicity and spectral types of metal poor subdwarfs. \cite{ryan91a,ryan91b} used metallic lines, such as the CaII\,K line to determine the metallicity of subdwarfs. The most widely used classification scheme for subdwarfs was introduced by \cite{reid95} and \cite{gizis97} who used CaH1, CaH2, CaH3, and TiO5 spectroscopic indices to classify subdwarfs. This system was later improved by \cite{lepine07} who devised the metallicity proxy $\zeta$, which is a third order polynomial of (CaH2+CaH3). The original $\zeta$ definition was recalibrated by \cite{dhital12}, who used wide binary pairs to improve the metallicity relation. \cite{lepine12} also introduced an improvement to $\zeta$, which is most effective for early type M and K dwarfs. \cite{lepine07} divided the subdwarfs into three metallicity subclasses in order of decreasing $\zeta$ -- subdwarfs (sdM), extreme subdwarfs (esdM), and ultra subdwarfs (usdM). These metallicity subclasses are characterized by decreasing TiO5 strength while the CaH remains relatively strong. 

An alternative system, without the use spectral indices, was devised by \cite{jao08}. The system was built upon the trend of the continuum from theoretical model spectra due to the complex dependence of the shape of the spectrum on temperature, metallicity, and gravity. They use M dwarf standard stars to derive the spectral subclass. In the \cite{jao08} system subdwarfs are not divided into ordinary, extreme, and ultra subdwarfs, but rather an independent metallicity strength is issued for each star.    

Since M subdwarfs represent one of the oldest stellar populations in the Galaxy, large sample statistics can provide invaluable information for Galactic kinematics and evolutionary history. For this purpose we need large numbers of subdwarfs, for which there is kinematic information. \cite{ryan91b} and \cite{gizis97} pointed out that the major kinematic difference between M dwarfs and subdwarfs is that the subdwarfs are part of the metal-poor Galactic halo (exhibiting little to no rotational motion), while dwarfs belong to the rotating disk population. Later, utilizing reduced proper motion diagrams and spectroscopic parallaxes, \cite{lepine03, lepine07} showed that subdwarfs are part of the halo population.

As members of the halo, subdwarfs are expected to have large Galactic velocities, and hence high proper motions. Traditionally, subdwarfs have been identified as large proper-motion stars in wide field surveys -- for example the Lowell Proper Motion Catalog \citep{carney94}, Luyten's LHS sample \citep{reid05a}, Lepine \& Shara Proper Motion catalog \citep[LSPM, ][]{lepine05}, or SuperCOSMOS \citep{subasavage05a,subasavage05b}. The current census of spectroscopically identified cool subdwarfs contains under 1000 stars and are included in the samples of \cite{hartwick84}, \cite{gizis97}, \cite{reid05},  \cite{lepine03,lepine07}, \cite{west04}, and \cite{jao08,jao11}. The SuperCOSMOS-RECONS group measured trigonometric parallaxes for about a hundred subdwarfs in a series of papers \citep{costa05, jao05, jao11}, determined absolute magnitudes and produced color-magnitude diagrams. However, most of the subdwarfs remain out of the reach of parallax studies -- absolute magnitudes and 3D kinematics have thus been challenging to determine. Recently, \cite{bochanski12} used the statistical parallax method to calibrate the subdwarf absolute magnitude scale. Absolute magnitudes estimates of subdwarfs permit the determination of reasonable distances to large samples of subwarfs that do not have measured trigonometric parallaxes. 

Previous studies have compiled samples of the most metal-rich classes of subdwarfs. A few tens of ultra subdwarfs have been identified in previous studies \citep{hartwick84,dawson88,ryanetal91,lepine07,burgasser07,lepine08}. \cite{jao08} included a number of very metal-poor subdwarfs in their sample, although they did not use the classification of extreme and ultra subdwarfs. Larger samples of all three kinds of subdwarfs, augmented by information about their distances, will prove invaluable for studies of Galactic kinematics and evolution.  

Recently, the physical properties of the M dwarf population have been extensively explored due to large photometric and spectroscopic samples \citep[e.g. ][]{kerber01, gizis02,reid05b,covey08, kowalski09, west11}. Large deep surveys such as the Sloan Digital Sky Survey \citep[SDSS;][]{york00} and the 2 Micron All Sky Survey \citep[2MASS;][]{skrutskie06} have proven efficient at building unprecedentedly large catalogs of cool stars \citep{reid08,west08,zhang09,kirkpatrick10,bochanski10,schmidt10,west11,folkes12}. Due to their low luminosities, the identification of large numbers of M dwarfs and subdwarfs has been extremely challenging and thus only stars within about 1-2\,kpc can be observed. The photometric and spectral capabilities of SDSS are specifically suited for faint stars, and thus the SDSS data are ideal for building large catalogs of cool subdwarfs. Even though SDSS possesses moderate spectral resolution (R$\sim$1800) it is sufficient for studying subdwarf spectra due the prominent spectral features in their spectra. \cite{west04} identified 60 subdwarfs from SDSS Data Release 2 (DR2) and \cite{lepine08} studied 23 cool ultra subdwarfs from DR6. \cite{west11} compiled a spectroscopic sample of about 70 000 M dwarfs as part of DR7 M dwarf spectroscopic catalog. Although \cite{west11} did not specifically identify subdwarfs from their spectra, this new sample contains  numerous subdwarf candidates based on the color ranges over which the spectroscopic sample was compiled.

In this paper we assemble a sample of 3517 spectroscopically identified cool subdwarfs from the SDSS DR7 sample of \cite{west11} that were not previously identified and a list of unidentified spectra removed during the creation of the DR7 M dwarf catalog \citep{west11} due to their ``odd'' spectral types. We use the new subdwarf catalog to study the statistical properties and kinematics of low-mass subdwarfs. In Section\,2 we discuss the observations and the selection criteria. In Section\,3 we present the radial velocity measurements and describe the process behind creating a new set of  spectral subdwarf templates. In Section\,4 we discuss the spectral and metallicity classification and the quality of the templates. In Section\,5 we examine the color-color and reduced proper motion diagrams in order to classify the bulk properties of subdwarfs of different metallicity classes. We discuss the $(U,V,W)$ Galactic velocity distributions and a fast-moving sample of stars in Section 6. In Section 7 we show color-magnitude diagrams. In Section 8 we identify a number of subdwarfs with H${\alpha}$ chromospheric activity and we discuss the possibility for intrinsic activity at these stellar metallicities and ages. We present a summary of the results and conclusions in Section 9.

\section{Observations}
For our analysis, we used data from the Seventh Data Release (DR7) of SDSS. The SDSS DR7 surveyed a region centered at the north Galactic cap and a smaller region in the South that covers\footnote{http://www.sdss.org/dr7/start/aboutdr7.html} $\sim$8\,423\,deg$^2$. The photometric data were collected in five filters ($u, g, r, i, z$) with photometric precision of 2\% at $r\lesssim 20$. The M dwarf spectroscopic candidates were chosen based on their colors, typical for cool stars: $0.5<r-i<3.05$ and $0.3<i-z<1.9$ from the DR7 photometric survey \citep{west04, west11}. Since we selected the subdwarf candidates from this sample, the same color criteria were applied. The standard SDSS pipeline produced spectra that were wavelength calibrated, sky subtracted, and have been shifted to the heliocentric rest frame. 

We assembled the final subdwarf catalog from two sources: 1) the DR7 cool stars catalog from \cite{west11}; and 2) a list of spectra of unidentified objects not included in the published catalog that were flagged as ``odd" during its construction. All SDSS DR7 spectra in the color ranges given above were examined by eye and all M dwarfs were included in the \cite{west11} catalog. In this process some unidentified or odd-looking spectra were separated from the sample. These unidentified spectra contained objects in the given color ranges, which were not identified as M dwarfs, but instead were binaries, other cool stars, galaxies, cataclysmic variables, etc. A sample of 400 candidate subdwarf spectra was assembled by eye from the initial list of 6,606 unidentified spectra. The visual identification was made based on comparisons of the target spectra to a collection of M dwarf template spectra from \cite{bochanski07}, and sample M subdwarf spectra from \cite{lepine07} and \cite{lepine08}. After removing misidentified M dwarfs, galaxies, binaries, and other cools stars from the unidentified spectra, this subsample consisted of 363 subdwarf candidates, which were later confirmed using the spectroscopic indices. 

The DR7 cool stars sample \citep{west11} had a record of the CaH and TiO indices and hence, the metallicity proxy $\zeta$ could be computed. We selected the M subdwarfs by requiring that $\zeta<0.825$ as specified in \cite{lepine07}, which resulted in 4,818 stars. Due to uncertainties in the spectra and the definition of $\zeta$, some M dwarfs leaked into the sample after the initial cut in $\zeta$. We visually inspected the entire sample and removed misidentified M dwarfs. From the original color-selected DR7 sample we identified 3,154 additional candidate subdwarfs that were not studied by \cite{west11}, although included in the catalog on the basis of their colors. The final sample that we present here contains a total of 3,517 subdwarfs (including the 363 stars from the ``odd"  spectra). This sample includes the 60 subdwarfs that were spectroscopically identified by \cite{west04}.

We used the SDSS DR7 web query\footnote{http://cas.sdss.org/astrodr7/en/tools/search/sql.asp} to extract the photometric equatorial coordinates, the $g,r,i$, and $z$ PSF magnitudes of all targets, and proper motions (in right ascension and declination) when available. The proper motions were determined based on a USNO-B match to SDSS \citep{munn04,munn08}. However, not all stars had measured proper motions due to the shallow red sensitivity of the USNO-B photometry. A total 2,368 stars in our subdwarf sample have measured proper motions. Our new sample increases the number of spectroscopically identified cool subdwarfs with proper motions by several times \citep[][]{gizis97,jao05,jao08,lepine07,lepine08,jao11,lodieu12,espinosa13}. A number of other parameters were computed based on previously determined values: $\zeta$, spectral and metallicity class \citep{lepine07,dhital12}, absolute magnitudes \citep{bochanski12}, distances (based on the distance modulus), height above the Galactic plane, Galactic velocities in the local standard of rest, and H$\alpha$ activity indicators \citep{west08,west11}. Most observed quantities are included in Table\,\ref{tbl-1}, and corresponding derived quantities are given in Table\,\ref{tbl-2}. All derived quantities will be discussed in the following sections. In addition to the parameters listed in Table\,\ref{tbl-1} and \,\ref{tbl-2}, we also recorded the uncertainty in the coordinates, photometric magnitudes, proper motions, a flag for the goodness of the proper motion \citep{munn04,munn08} as well as errors in the derived quantities, such as distance and tangential velocity. The complete catalog is available as online material to this paper.

\section{Radial Velocities and Template Assembly}
The spectral classification of subdwarfs is an important step in identifying their association with different metallicity and spectral types. Together with the radial velocities (RVs) and distances, spectral and metallicity classes are the first parameters needed to derive statistical information about the kinematics and bulk properties of subdwarfs. A precise way to determine radial velocities is by cross-correlating the target spectrum with a rest-frame template spectrum \citep{td79}. Spectral classification ensures precise template matches before cross-correlation is carried out. The subdwarfs from the DR7 cool stars sample all had previously measured RVs based on cross-correlation with M dwarf spectra from \cite{bochanski07}, which might not be entirely accurate due to the higher metallicity of the templates and the potential spectral type mismatch. One of our motivations for creating a subdwarf catalog is the production of subdwarf template spectra for each spectral subclass of all three metallicity classes in order to determine more precise radial velocities.

For this purpose, we first determined the wavelength shifts by fitting single Gaussian profiles to a set of prominent absorption lines of neutral metals for all 363 spectra from the ``odd'' list. We used seven different lines, which are listed in Table\,\ref{tbl-3} along with their rest vacuum wavelengths, obtained from the spectral line database of the National Institute of Standards and Technology (NIST\footnote{http://physics.nist.gov/PhysRefData/ASD/lines\_form.html}). The final wavelength shift was determined based on a weighted average of the shifts from all detected lines. Most spectra did not have good enough S/N in all seven lines, which was taken into account either by excluding some lines that did not have good quality, excluding outliers from the distribution of Gaussian means, or down-weighting some lines based on the uncertainty in the mean. The uncertainty in the RV was determined based on the standard deviation of the velocities measured from at least three lines. The instrument resolution is the major source of uncertainty in these calculation. The typical 1$\sigma$ RV uncertainty is of the order of 10-20\,km\,s$^{-1}$. The fraction of the stars with velocity uncertainty less than 10\,km/s is 58\%.

The assembly of the template spectra was a two-step process: First, the RVs for the subdwarfs from the ``odd'' sample were used to return all ``odd'' spectra the to the rest frame so that the subdwarfs could be spectral typed and classified (see Section\,4). The spectra with S/N$>5$ in both the CaH and TiO features from this smaller sample were selected to create preliminary template spectra for all integer spectral subclasses and for each of the metallicity classes. All spectra used to make templates were spline interpolated to 15 times higher resolution as compared to the original one. This is justified by the fact that we can obtain radial velocity precision better than a resolution element \citep{bochanski07}. Therefore, we subgrided the template spectra, co-added them and corrected for the radial velocity, which yielded a more precise and higher resolution template. Templates were assembled by first normalizing the flux in all spectra to the value at 7500\AA\, and then computing the mean of the spectra belonging to the same subclass. 

These templates were then used to determine the RVs of all 3517 stars based on a cross-correlation technique.  A histogram of the final RVs can be found in Figure\,\ref{rvhist}. There are a large number of stars with large RVs, which is consistent with subdwarfs being members of the dynamically heated older disk or halo population. After all spectra were returned to the rest frame and spectral typed (see Section\,4), we applied the same procedure described above to assemble the final set of templates. We discuss the features of the templates in the next section. We provide these templates as online material to this paper.

\section{Spectral Typing and Template Features}

There are two major classification schemes for low-mass subdwarfs, outlined in \cite{lepine07} and \cite{jao08}. The L{\'e}pine system is entirely empirical based on the relative strengths of the spectral indices, while the Jao system relies on an overall match of the target spectrum with model spectra of M dwarfs. In the Jao system, both spectral features and continuum are matched to the model spectra, looking at the overall appearance of the spectrum as a whole. The Jao system provides a very detailed consideration of the effect of temperature, metallicity, and gravity on the different spectral features. However, for the purpose of this study we employed the L{\'e}pine system based on the spectral indices, which is more manageable for large numbers of subdwarf candidates and allows us to directly compare to several previous subdwarf investigations. We leave the comparison of the Jao and L{\'e}pine systems using our large sample of subdwarfs for a future study. It may be possible to arrive at a subdwarf classification system that employs the best of both systems. 

As mentioned in Section 1, the \cite{lepine07} system was improved to better match two different populations of common proper motion stars -- early type K and M dwarfs \citep{lepine12}, and high galactic latitude early-type M dwarf-white dwarf binaries \citep{dhital12}. Although both recalibrations of $\zeta$ are not focused on the subdwarf region in (CaH2+CaH3) vs. TiO5 space, we take the word of caution from \cite{lepine12}, that the different calibrations depend on the raw observations and with which observatory they were obtained. Thus, for our analysis we adopted the treatment of $\zeta$ given in \cite{dhital12} since the binaries they used for the recalibration and our subdwarfs were both observed with SDSS. 

The strength of the TiO5, CaH2, and CaH3 band heads were measured according to the L{\'e}pine spectral classification systems. We measured the strength of these features and the corresponding continua in the wavelength ranges given in Table\,\ref{tbl-4}, adopted from \cite{gizis97}. In this study, we employed the formalism of \cite{lepine07} for determining metallicity classes and spectral subclasses based on the relative strengths of the TiO5 and the CaH band heads. The metallicity proxy $\zeta$ was computed by comparing the target metallicity to the solar value for given CaH and TiO indices \citep{lepine07,dhital12}. We used the same dividers in CaH-TiO space as in \cite{lepine07} to determine the metallicity classes: $0.5<\zeta<0.825$ for sdMs, $0.2<\zeta<0.5$ for esdMs, and $\zeta<0.2$ for usdMs. The numerical spectral subclass is defined by a polynomial fit to the combined strength of the CaH2 and CaH3 bands as given in equation (3) in \cite{lepine07}: 
\begin{equation}
\mathrm{SpT}=1.4(\mathrm{CaH2+CaH3})^2-10.0(\mathrm{CaH2+CaH3})+1.24
\end{equation}
The result was rounded to the nearest integer. We used this expression for the entire range of metallicities and temperatures since it has already been applied to such stars extensively by \cite{lepine07}.

All stars of the final catalog were separated in metallicity classes and spectral subclasses. The distributions of the resulting spectral and metallicity types are shown in Figure\,\ref{spstat}. The majority of the stars are in the sdM class (2078), with fewer in the esdM (905), and usdM (534) classes. While the sdMs peak in spectral subclass 3, the esdMs and usdMs peak at 1. As expected, we have very few ultracool stars with spectral subtypes 5 or greater. In fact, we identified only 26 ultracool stars that might be of further interest as low-temperature, low-metallicity objects. Figure\,\ref{ind} shows the sample plotted in CaH2+CaH3-TiO5 space \citep{lepine05, lepine07}. The boundaries between the different subdwarf classes are clearly seen.

As mentioned in the previous section, we have used this spectral typing to build spectral templates for radial velocity determinations. Figure\,\ref{3temp} demonstrates the differences among the template spectra for stars of the same spectral subclass (M3) of the three metallicity classes of subdwarfs. The locations of the prominent lines and bandheads are marked on the top most spectrum of Figure\,\ref{3temp}. The metallicity effects dominate in the grey shaded regions; the depth of the TiO feature decreases with increasing metallicity, which also changes the wings of the K\,I line. However, we do not clearly see metallicity effects in the 6340\AA\,-6500\AA\,\ and 6500\AA\,-6900\AA\,\ spectral ranges, which also contain TiO features (cyan shaded regions). The cyan shaded regions most probably also have temperature effect. We also do not see such metallicity effects in these regions in the example spectra from \cite{lepine07}.  

Template spectra for the sdM class are shown in Figure\,\ref{temp1}, and Figures \ref{temp2} and \ref{temp3} show the templates for esdM and usdM metallicity classes respectively. The effect of metallicity in the spectra can be clearly seen: the TiO band is reduced, while the CaH band remains strong as metallicity is decreased. However, the spectra for different spectral subtypes of the same metallicity class look somewhat similar. \cite{bochanski07} and \cite{kirkpatrick92} discus that the temperature difference for M dwarf spectra between M0 and M9 is 1400K (from 3800K to 2400K), and a temperature effect is seen in M dwarf spectra in the TiO bandheads at 7126\AA\,-7135\AA\,\ and 7666\AA\,-7861\AA. \cite{wing76} and \cite{reid95} mention that these TiO features are both metallicity and temperature dependent. However, our spectra do not show such an obvious temperature dependence. In the L{\'e}pine system, TiO is taken to be mainly metallicity dependent: the example spectra given in \cite{lepine07} show very little temperature dependence of the TiO bandhead. The \cite{jao08} system, which takes both the temperature and metallicity effects into account, might be able to resolve this issue. For example, \cite{jao08} proposes that the slope of the continuum between 8200\AA\,\ and 9000\AA\,\ be used to determine the temperature-dependent subclass, and the TiO features at 7050\AA\,-7150\AA\,\ be used for metallicity classification. The effects of metallicity and temperature in the M dwarf spectra are seen in Figure\,9 of \cite{jao08}. In the future we will recompute our templates in this system and investigate how it compares to the L{\'e}pine classification. As a first step we compare several of our templates for different metallicity classes and spectral subtypes (using the L{\'e}pine system) with some of the spectra from \cite{jao08} (classified in the Jao system). It is clear from Figure\,\ref{tempjao} that the two systems have significant differences in the early spectral classes -- 0 and 1 for all metallicity classes. The differences are as expected from the above analysis - in the green-shaded regions from Figure\,\ref{3temp} and the region of the KI doublet. The spectra for spectral classes 3 and 5 look very similar. In addition, \cite{jao08} suggest that the sdM, esdM, usdM categories may not represent true metallicity subclasses. These issues will be resolved when accurate Fe/H scalings based on optical spectra are available \citep{mann13}. There are potential issues with either classification system but we are confident that we have identified a robust sample of of low-mass subdwarfs. We leave further discussion of the advantages and disadvantages of both to a future study.

\section{Color-color and Reduced Proper Motion Diagrams}
\subsection{Colors}
Previous studies have found that subdwarfs separate from M dwarfs in color-color diagrams \citep{west04,lepine07}. Figure\,\ref{ccd} shows a color-color diagram ($r-z,g-r$) for all classified subdwarfs in the current catalog with measured proper motions. The different metallicity subclasses are shown as different colors, similar to the previous figures (purple for sdMs, green for esdMs, and red for usdMs), and the distribution of field M dwarfs with high proper motions ($\mu>30$\,mas/yr) from the DR7 catalog is shown as small blue dots. We chose to display only high proper motion field M dwarfs since such comparisons between M dwarfs and subdwarfs have been done before only for high proper motion stars and thus this aids the comparison with previous studies \citep[e.g.][]{lepine07}.

Because our SDSS sample of subdwarfs is unprecedented in its large size, particularly in the numbers of esdMs and usdMs, we can explore how subdwarfs separate in color-color space. Figure\,\ref{ccd} shows the clear segregation of the subdwarfs and field M dwarfs in $g-r$ vs. $r-z$ color. There is also a slight separation between the three metallicity classes of subdwarfs. It is evident that the subdwarfs lie in the expected range of colors for low-mass dwarfs, but have systematically redder $g-r$ colors than the field dMs by about 0.2 magnitudes \citep[noted by previous studies; e.g.][]{west04}. \cite{lepine07} showed a similar result with respect to the $V-J$ color. From our sample, we see that 82\% of the stars with $g-r$ $>$ 1.6 are classified as subdwarfs. In addition to separation from the M dwarfs, the different metallicity classes are also somewhat separated from each other although there is some overlap: sdMs are about 0.2\,mag redder in $g-r$ than M dwarfs, esdMs -- 0.3\,mag, and usdMs -- 0.4\,mag redder. Table\,\ref{tbl-5} gives the mean $g-r$ and $r-z$ colors for all metallicity classes and the M dwarfs plotted in Figure\,\ref{ccd}. In Table\,\ref{tbl-5}, we have separated the stars by spectral class and the mean colors in bins of two spectral classes because the individual spectral classes do not have enough stars for precise color statistics. There is a slight trend of increasing $r-z$ and $g-r$ with spectral class. While in general, the subdwarfs are best identified by their redder $g-r$ colors with a successful rate of 82\%, the separation between the three subdwarf classes is most prominent in the $r-z$ color. There are some regions of color-color space where the different metallicity classes are well separated but there is still some considerable overlap. Thus, while color-color diagrams might be sufficient to separate subdwarfs from disk dwarfs, they do not appear to be able to clearly separate the metallicity subclasses in the overlap regions. Spectroscopic data using the CaH-TiO molecular bands appear to be required for be a robust subdwarf classification.

\subsection{Reduced Proper Motion Diagrams}
Another method which has been traditionally used to efficiently separate different populations of stars is the use of reduced proper motion \citep[RPM, ][]{luyten22} diagrams in visual and infrared colors \citep{subasavage05b,lepine07,faherty09}. This technique is motivated by the fact that RPM diagrams provide information about the kinematic classification of large samples of stars when the stars lack measured distances, since their characteristic Galactic velocities are reflected in the typical values of their proper motions and apparent magnitudes. The reduced proper motion is defined as:
\begin{eqnarray}
H_r=r+5\log{\mu}+5\\
H_r=M_r+5\log{\upsilon_{\mathrm{T}}}-3.38,
\end{eqnarray}  
where $H_{\mathrm{r}}$ is the reduced proper motion in the SDSS $r$ band, $\mu$ is the total proper motion in arcseconds per year, $M_{\mathrm{r}}$ is the absolute magnitude in $r$, and $\upsilon_{\mathrm{T}}$ is the transverse velocity in km\,s$^{-1}$. From (2) we see that the reduced proper motion diagram (reduced proper motion versus color) is similar to a color-magnitude diagram. Equation (3) shows the connection between RPM, the absolute magnitude, and the transverse velocity. The transverse velocity complicates a direct correspondence between $H_{\mathrm{r}}$ and $M_{\mathrm{r}}$ causing a larger spread in $H_{\mathrm{ r}}$-color diagrams as compared to absolute magnitude-color diagrams with absolute magnitudes derived from trigonometric parallax. Not all stars have the same Galactic motions, which introduces scatter in $\upsilon_T$ and creates a spread in RPM. Since $\mu$ is a directly observable quantity for many nearby stars, the RPM diagrams do not suffer from the additional errors accumulated when determining the absolute magnitude based on fits to the color or spectral type.

Figure\,\ref{ccrpm} shows an RPM diagram in $g-r$ color. The colors of the dots are the same as in previous figures. The different metallicity classes are plotted separately and we see that they are slightly segregated in reduced proper motion space. The bulk of the sdMs and field M dwarfs are shifted in color with respect to each other but their $H_{\mathrm{r}}$ range is similar, implying that they most probably belong to a similar kinematic population -- a Galactic disk population. On the other hand, the esdMs and usdMs populations do not differ significantly on the diagram, but both have preferentially larger values of RPM than the sdMs and dMs. This separation can also be seen from the mean and standard deviation of the RPMs for the three metallicity classes and in bins of spectral class given in Table\,\ref{tbl-hr}. As discussed in \cite{lepine05}, the larger RPM values are indicative of halo population objects. This difference can be explained by the chemical evolution of the Milky Way: lowest metallicity stars belong to an older population, which has been dynamically heated for a longer time or are members of accreted systems, and hence have acquired larger velocities and orbits typical of the halo population. As can be seen from Table\,\ref{tbl-hr} the spread is large in the different spectral classes and the shift in values to progressively larger RPM with spectral class is generally within the one sigma spread.  

To show that stars with large radial and tangential velocities lie in the high-RPM domain, Figure\,\ref{rvrpm}(top) plots stars with large radial velocities (RV$>$150\,km\,s$^{-1}$) shown as green dots and the remainder of the stars -- as black triangles. The stars with large radial velocities cluster at high values of $H_{\mathrm{r}}$, as is expected for halo population stars. The field M-dwarf population is also plotted for reference. In the bottom part of Figure\,\ref{rvrpm}, stars with large tangential velocities ($\upsilon_{\mathrm{T}}>$200\,km\,s$^{-1}$) are plotted. The process of calculating the transverse velocity is discussed in the Section 6. The bottom of Figure\,\ref{rvrpm} simply shows the effect of Eq.\,(3), where the dependence of $H_{\mathrm{r}}$ on $\upsilon_{\mathrm{T}}$ is shown. Figure\,\ref{rvrpm} demonstrates that the stars with large radial velocities overlap to some extent with the stars with high tangential velocities, and they are both situated in the part of the the RPM diagram occupied by esdMs and usdMs that have preferentially high RPMs.

\section{Kinematics}
\subsection{Distances}
As mentioned above, it is instructive to use the subdwarf sample to infer information about Galactic kinematics. Traditionally, subdwarfs were thought to be part of the halo population \citep{ryan91b,gizis97,lepine03,lepine07} and more rarely -- from the tick disk \citep{monteiro06}. However, as seen in the analysis of the RPM diagrams above, it is probable that only the esdM and usdM metallicity classes of subdwarfs belong to the halo. The sdMs are less dynamically heated and possibly belong to the old Galactic or ``thick"  disk. Examining the three-dimensional Galactic space motions allows us to probe this conclusion in further detail. To obtain the 3D kinematics, we first had to estimate the distances to the objects in the sample. 

We determined the distances from the distance modulus, using the absolute magnitudes-color relations from \cite{bochanski12}, visual magnitudes and correcting for extinction. The absolute magnitudes, $M_r$, $M_i$ and $M_z$, are determined employing the formalism of \cite{bochanski12}, who used the maximum likelihood formulation of the statistical parallax analysis \citep{murray83, hawley86} on our SDSS sample of subdwarfs. The result of this analysis produces the kinematics of the selected groups of stars relative to the Sun and their absolute magnitudes \citep{bochanski12}. The typical error of the statistical parallax for subdwarfs is (0.1-0.4) magnitudes, as compared to trigonometric parallax which can give magnitude uncertainty as little as 0.02 magnitudes \citep{koen92}. From the computed absolute magnitudes in the $r$-band, we calculated the distances. Distances computed in the $i$ or $z$ band may not necessarily agree since the three absolute magnitudes have been computed independently from \cite{bochanski12} with no prior knowledge of the distance. The extinction in SDSS is determined based on the extinction maps produced by \cite{schlegel98} and the numbers given by the SDSS query account for the entire line of sight, which for nearby subdwarfs may be a slight overestimate. Therefore, the extinction correction did not have a large effect on our results since typical values of the extinction are small (0.05-0.1 magnitudes) in the redder bands, and all values used can be found in Table\,\ref{tbl-1}. The uncertainty in the photometric visual magnitudes is also small, making the dominant source of error in the distance that of the absolute magnitude. The typical uncertainty in the distance coming from the uncertainty in the absolute and apparent magnitudes, and extinction is 20\%. A histogram of the derived distances is shown in Figure\,\ref{dist}. Since subdwarfs are intrinsically faint, Figure\,\ref{dist} shows that we can only see stars within about 1.5\,kpc of the Sun, with most of the stars located at distances of about 300-400\,pc. We identified eight subdwarfs with distances within 50\,pc from the Sun. 

After we computed the distances to the stars, we estimated their tangential velocities ($\upsilon_{\mathrm{T}}$) based on their proper motions and distances. The typical error in determining $\upsilon_{\mathrm{T}}$ is 30\,km\,s$^{-1}$. Figure\,\ref{vthist} shows the distribution of transverse velocities. There is a significant tail of the distribution towards high tangential velocities. The tangential velocities and their uncertainties are included in the catalog.

Based on the distance and the galactic coordinates of the stars we determined the distance from the Galactic plane (also included in the catalog) based on the value for the Sun of 15\,pc above the Galactic plane and 8.5\,kpc from the Galactic center \citep{majaess09}. This vertical distance from the plane has been used as a proxy for age in a number of studies \citep[e.g.][]{west04,west08}. Figure\,\ref{zeta} shows how the mean and standard deviation (dashed error bars) of the metallicity proxy $\zeta$ vary with distance from the Galactic plane for all subdwarfs in the sample. $\zeta$ shows a slight trend of decreasing from above 0.5 (the dividing value between sdMs and esdMs), in the first four bins to under this value at 450\,pc away from the plane within the error in the mean (solid) error bars. The error in the mean is computed for simplicity as the the standard deviation divided by the square root of the number of stars in each bin. This assumes that standard deviations in each measurement are the same.{ This is not necessarily the case because the stars have different S/N, so these error bars might be slightly underestimated.} The standard deviation in $\zeta$ shows that the scatter of the stars in each distance bin is larger close to the disk and gets smaller as the distance from the Galactic plane increases; small distances have a broader mixture of subdwarfs with different metallicities, while farther away we find mostly subdwarfs with low metallicities. The bin at 900\,pc deviates from this trend, but it contains the smallest number of stars and is not statistically significant.

\subsection{3D Galactic Motions}
The final portion of the kinematic analysis comes from the determination of the 3D space velocities. Several Galactic population studies have already computed average velocities and velocity dispersions characteristic for the different structural components of the Galaxy \citep[e.g.][]{casertano90,ivezic08}. We determined the 3D space velocities in the classic galactic system ($U,V,W$) using the distances, radial velocities, proper motions, and the coordinates of the stars. All velocities were calculated in the local standard of rest, corrected for the solar motion assuming $(U, V, W)=$[-11.1, 12.24, 7.25]\,km\,s$^{-1}$ for the Sun \citep{sch10}. In this system, stars from the thin disk (0-100\,pc) have $\langle V\rangle=-20$\,km\,s$^{-1}$, and velocity dispersions $\sigma_U=32$\,km\,s$^{-1}$, $\sigma_V=23$\,km\,s$^{-1}$, $\sigma_W=18$\,km\,s$^{-1}$ \citep{fuchs09}. The thick disk (700-800\,pc) has $\langle V\rangle=-32$\,km\,s$^{-1}$, and $\sigma_U=49$\,km\,s$^{-1}$, $\sigma_V=38$\,km\,s$^{-1}$, $\sigma_W=40$\,km\,s$^{-1}$ \citep{fuchs09}, and the typical values for the halo ($<$25\,kpc) are $\langle V\rangle=-173$\,km\,s$^{-1}$, and $\sigma_U=135$\,km\,s$^{-1}$, $\sigma_V=105$\,km\,s$^{-1}$, $\sigma_W=90$\,km\,s$^{-1}$ \citep{binney98}. These values are in the direction of the Sun's motion in the LSR frame.

Figure\,\ref{vu} shows a plot of the $U$ versus $V$ velocities for the four metallicity classes of stars, represented in the same colors as in earlier figures. While the field M dwarfs (blue contours) are centered at (0,0), the  metal poor stars extend to more negative $V$ velocities, while still remaining at $U=0$\,km\,s$^{-1}$ (all corresponding contours are at { the 40\%, 68\% and 95\% levels}). The extent of the contours in $V$ of the sdMs (purple) is certainly less than that of the esdMs (green) and usdMs (red), which have a similar shape. We performed a Kolmogorov-Smirnoff test to determine whether any two of the distributions were drawn from the same parent distribution. All of the probabilities were of the order of $10^{-5}$ except the probability that esdMs and usdMs were drawn from the same population (0.32), implying that kinematically the two most metal poor classes of subdwarfs have similar kinematics at statistically significant level. 

A more detailed analysis of the 3D velocities is presented in Figure\,\ref{uvwhist}, where distributions of the three velocity components are shown with the y-axis in logarithmic scale to show the differences in the distributions more clearly. Both the $U$ and $W$ velocities are centered at zero for all four metallicity classes, with the sdMs (purple) and esdMs (green) having a broader tail in $U$ with dispersion of 101\,km\,s$^{-1}$. The biggest difference in the distributions is in the $V$-component. This difference is expected for different stellar populations due to the asymmetric drift of stellar orbits. The field M dwarfs have the smallest tail in $V$ reaching to -400\,km\,s$^{-1}$ and peak at around 0\,km\,s$^{-1}$, while the esdMs (green) and usdMs (red) peak at highly negative relative velocities (-155\,km\,s$^{-1}$ and -171\,km\,s$^{-1}$ respectively). The means and velocity dispersions for the three subdwarf classes are given in Table\,\ref{tbl-6}. We computed the mean and standard deviations based on Gaussian distributions, which is a much better assumption for the esdMs and usdMs, than for the sdMs and the field M dwarfs, which display significantly skewed distributions. These values agree well with the values derived in \cite{bochanski12}, who used the method of statistical parallax to make a similar determination. Comparing these numbers with the characteristic values for the different galactic components, we infer that esdMs and usdMs belong to the halo population, while the ordinary subdwarfs belong to the thick or old thin disk.{ The mean $V$ velocities for all metallicity classes suggest that these stars have experienced strong asymmetric drift as a consequence of dynamical heating and their orbits are strongly elliptical. In addition, the large dispersions in all three components, especially for the esdMs and usdMs suggest that their orbits do not lie in the disk and hence are members of the halo}. As can be seen here from our detailed analysis of a large sample of subdwarfs, only the most metal poor subdwarfs can be attributed to the halo. In fact, the majority of the subdwarfs by number in our catalog likely belong to the thick or old stellar disk.      

Having calculated the three components of the galactic motion of the subdwarfs in our sample, we could calculate the total velocity, summing them in quadrature. We find 14 stars with S/N$>$5 in the TiO5 spectral feature that have total velocities more than 525\,km\,s$^{-1}$, which we take to be the Galactic escape velocity in the solar neighborhood \citep{carney87}. These fast stars are interesting because they will eventually escape the Milky Way. We list the parameters of these stars in Table\,\ref{tbl-7}. A future study will contain more analysis on these high-velocity stars.

\section{Color-Magnitude Diagrams}

Using the absolute magnitudes from \cite{bochanski12}, we produced color-absolute magnitude diagrams for a large sample of subdwarfs. The only information on the absolute magnitudes of subdwafs in previous studies was available from parallax measurements of about a hundred of bright stars total \citep{gizis97,subasavage05b,jao05,costa05}. These authors showed color-magnitude diagrams (CMD) for such stars demonstrating that subdwarfs lie under (or to the left of) the main sequence of other bright stars with good parallax measurements. 

Figure\,\ref{Mr-zr} shows a CMD -- $M_r$ vs. $r-z$ color (top). The main sequence of the field M dwarfs is shown in blue. The absolute magnitudes of the M dwarfs were calculated using the polynomials given in \cite{bochanski10}. It is clear that most subdwarfs lie under (or to the left of) the main sequence (MS) represented by the metal-rich M dwarfs. The esdMs lie farther from the MS and usdMs even farther showing less scatter in the absolute magnitudes. All subdwarfs overlap in the region $[M_{\mathrm{r}},r-z]\sim[11-13,0.9-1.3]$. A tighter distribution of subdwarfs is shown in another version of CMD -- $M_r$ vs. the $g-r$ color (Figure\,\ref{Mr-zr} bottom). The edge of the subdwarf sequence is clear and sharp, because $M_r$ is a fitted relation as a function of $r-i$ color and we select our sample by this color. When the field M dwarfs are plotted on such a diagram they do not show a clearly defined MS, and have thus been omitted. Unfortunately, $M_g$ has not been computed for the field M dwarfs \citep{bochanski10} nor for the subdwarfs \citep{bochanski12}. This tight relationship is expected since the distribution in $g-r$ color is significantly more confined to the area close to the MS than the $r-z$ sequence as can be seen from Fig.\,\ref{ccd}. However, the lack of separation of the different metallicity classes in these color-magnitude diagrams is somewhat counterintuitive, i.e. we might expect that the more metal-poor the stars the farther away from the MS they lie. It is possible that a metallicity effect, such as the presence of strong features like MgH and CaH in this color bands is changing the slope of the spectrum from blue to red, thus shifting the more metal poor stars to the right (redder colors) and hence closer to the MS, although they still lie under or to the left of it.       

\section{Magnetic Activity}

When assembling the sample, we noticed that some spectra showed a significant H$\alpha$ line emission, which is a strong indicator of chromospheric activity \citep{hawley96, west04}. A sample spectrum of an active sdM3 star is shown along with an inactive sdM3 in Figure\,\ref{act}. We measured the equivalent width (EW) of the H$\alpha$ line and used the four criteria for  designating a star as active defined by \cite{west11}: (1) H$\alpha$ EW must be greater than 0.75\,\AA; (2) The uncertainty in the EW must be less than the value; (3) S/N$>$3 at the H$\alpha$ line; (4) The height of the H$\alpha$ line must be 3 times greater than the nearby ``continuum'' region. In this sense, if a star fulfills all four criteria it is assigned an activity flag of 1. A star is assigned 0 if inactive when the third criterion is fulfilled and the rest of the criteria fail, or a -9999 if the third criterion is not fulfilled. The values of the activity flag along with the H$\alpha$ equivalent widths and error estimates of the EW are all recorded in our catalog (see Table\,\ref{tbl-2}). Following these criteria we found total of 208 active stars from the entire sample - 134 sdMs, 41 esdMs, 33 usdMs. A histogram of the active stars in different spectral subclasses for the three metallicity classes of sudwarfs is shown in Figure\,\ref{actsp}. Most of the active sdMs are in spectral subclass 3, while the esdMs and usdMs show activity in earlier spectral classes. Chromospheric activity in low-mass subdwarfs has only been reported once before by \cite{west04}, but this is the first statistical study of active low-mass subdwarfs. In Figure\,\ref{actspbin}, we show the activity fractions of all subdwarfs (upper left) and three metallicity classes in bins of spectral subtype. While the sdMs show a clear rise and then a level-off of activity with spectral class, such an effect is not seen for the esdMs or usdMs, which are basically flat within the error bars. That also suggests that in terms of activity ordinary subdwarfs behave differently than the their more metal-poor counterparts, which might simply be an effect of the distance from the Galactic plane (or age). In fact, the distribution of activity fraction for sdMs looks remarkably similar to the results from activity studies of normal disk dwarfs \citep{west08}.

There are two possibilities to explain these activity fractions - intrinsic activity on the subdwarfs themselves or subdwarfs in tight binary systems. Recently \cite{morgan12} used a large sample of close white dwarf-M dwarf pairs to demonstrate statistically that the presence of a close companion prolongs the active phase of M dwarfs. Because magnetic activity is best studied in Galactic context, we investigated how the activity fraction varies in bins of distance from the Galactic plane (Figure\,\ref{actd}). In the upper left panel of Figure\,\ref{actd} the activity fractions for all subdwarfs in the sample are plotted. There is a trend of the activity fraction decreasing with increasing distance from the Galactic plane for all metallicity classes; the effect is strongest when we combine all sdMs. The error bars shown are one sigma, assigned based on binomial statistics and it is evident that within the error bars this decreasing trend is statistically significant. This can be explained as an age effect due to dynamical heating; old stars have perturbed orbits that get farther from the Galactic disk \citep{west06, west08}. If activity is dependent on age (and the distance from the Galactic plane is correlated with age), then the observed distributions of subdwarfs can be explained as having finite active lifetime. Thus, stars farther from the Galactic plane are older and turned off \citep[see][for more details]{west08}. Figure\,\ref{actd} compares well with Figure\,5 of \cite{west08} of the distribution of later type M dwarfs. However, this does not rule out enhanced activity from the presence of a companion. The same fall-off with distance from the Galactic plane exists in the activity fractions for tight M dwarf -- white dwarf binaries, indicating that while close pairs prolong the active lifetimes, they are still finite \citep{morgan12}. We therefore cannot say with certainty that the observed effect indicates intrinsic activity on subdwarfs or is a result of a close companion. What is particularly interesting is that a decrease of activity with Galactic height is expected if sdMs are old disk members, but is not necessarily expected for esdMs and usdMs distributions. These questions can be resolved by employing high S/N spectra, complete high-resolution photometric observations, and/or obtaining precise trigonometric parallax measurements to a large sample of subdwarfs, which can shed light on subdwarf multiplicity. Some initial steps towards these goals have been completed in recent years \citep[e.g.][]{jao11}.

In addition, we examine the ratio between the luminosity in H$\alpha$ ($L_{\mathrm{H\alpha}}$) and the bolometric luminosities ($L_{\mathrm{bol}}$) of the stars. This is done because the EW depends on the neighboring continuum, which is different for stars of different spectral classes, and the $L_{\mathrm{H\alpha}}/L_{\mathrm{bol}}$ is an independent measure. $L_{\mathrm{H\alpha}}/L_{\mathrm{bol}}$ is calculated using the method of \cite{west04} and \cite{walkowicz04} for M dwarfs. We used the $\chi=L\mathrm{(cont)_{H\alpha}}/L_{\mathrm{bol}}$ given as a function of SDSS $r-i$ color from \cite{walkowicz04}. Figure\,\ref{lha} shows no clear trend with distance from the Galactic plane. In Figure\,\ref{lha-sp} we show the mean value of $L_{\mathrm{H\alpha}}/L_{\mathrm{bol}}$ as a function of spectral class for the three metallicity classes of subdwarfs. It is evident that for the sdMs the activity measure { slightly decreases} with spectral class within the solid error bars, which may be due { to systematically lower $L_{\mathrm{H\alpha}}/L_{\mathrm{bol}}$ with good S/N with spectral class. There is no clear trend for the other metallicity classes. The solid error bars in Figures\,\,\ref{lha} and \ref{lha-sp} are both computed by propagating the individual errors on the quantities in the expression for $L_{\mathrm{H\alpha}}/L_{\mathrm{bol}}$ to obtain the uncertainty in each  $L_{\mathrm{H\alpha}}/L_{\mathrm{bol}}$. Then taking these individual uncertainties into account we compute the error in the mean as in \cite{westH08}. The data for all subdwarfs and sdMs} follow a similar distribution to Figure\,5 of \cite{west04}, hinting on the similarity between subdwarfs and field M dwarfs.

\section{Summary and Conclusions}

We present the largest single sample of cool subdwarfs compiled from the 7th data release of the Sloan Digital Sky Survey. The complete catalog contains 3,517 subdwarfs, out of which 2368 have measured proper motions. The catalog is provided in the online materials accompanying this paper. Our catalog consists of all directly measured quantities such as photometric magnitudes, proper motions, and astrometry, as well as a comprehensive list of derived quantities such as absolute magnitudes, distances, activity, 3D galactic velocities. This catalog significantly increases the total number of spectroscopically identified subdwarfs in previous studies. In this work, we put forward some of the possible analysis that can be done with such a large sample of cool nearby subdwarfs. Here we present the summary of the presented work and main results.

We spectral typed and classified all stars into three metallicity classes as suggested by \cite{lepine07} and performed statistical analyses on the stars in the various metallicity classes. We acknowledge that there might be some potential issues with the current \cite{lepine07} classification system. Future analysis will contain a more detailed comparison between the \cite{lepine07} and \cite{jao08} systems using our large sample of subdwarfs.

We examine color-color and reduced proper motion (RPM) diagrams, where we show a clear segregation in color and RPM of the subdwarfs from field M dwarfs, and sdMs from esdMs and usdMs, but virtually no separation between esdMs and usdMs. The RPM diagrams combined with the knowledge of 3D space motions in the ($U, V, W$) system indicates that while the most metal rich of the subdwarfs, the sdMs, belong to the thick disk (or old disk) population in the Galaxy, the two more metal-poor classes, esdMs and usdMs, likely belong to the halo. Traditionally, all subdwarfs have been thought to belong to the halo \citep[e.g.][]{lepine07}, but here we show that different metallicity classes have different dynamics, with the esdMs and usdMs being more dynamically heated. This effect is well explained if esdMs and usdMs are older than sdMs, having the lowest metallicity and thus tracing an older period in the star formation history of the Galaxy. This is further evidenced by the fact that the activity fraction decreases as the distance from the Galactic plane increases, which can be used as a proxy of age \citep{west06, west08, west11}. The subdwarf kinematics allowed us to determine the absolute magnitudes of a large sample of subdwarfs \citep{bochanski12}, which gave us the opportunity to produce color-magnitude diagrams, where the segregation of subdwarfs and field M dwarfs can be clearly seen. Again, the esdMs and usdMs do not separate significantly in the CMDs. We report that esdMs and usdMs are more confined to the main sequence while the sdMs show a significant spread when viewed in $g-r$ color, and the opposite effect in $r-z$ color. All subdwarfs form a metal-poor sequence of stars that lies under (or to the left of) the main sequence of field M dwarfs as expected. 

Additionally, we find that 6\% of the subdwarfs in our sample exhibit magnetic activity and the number of active stars is highest in the sdM class and falls off with metallicity, which partly traces the total number of stars in the three different metallicity classes. We show that the activity fraction for all metallicity classes falls of with distance from the Galactic plane and increases toward later spectral class similar to what is seen in field M dwarfs. The puzzle of whether subdwarfs are intrinsically active or members of binary systems containing active companions can be explored further in more detail with high resolution photometry and spectroscopy. We also study the luminosity in H$\alpha$ ($L_{\mathrm{H\alpha}}/L_{\mathrm{bol}}$ ) and show a slight trend to { decrease} with spectral class.

In addition, after we compute the 3D space motions we identify a list of 14 fast moving stars that are moving with total velocity larger than the escape velocity of the Galaxy. We leave the study of the fast members and possible co-moving groups of stars for a future study. The latter can be indicative of the past merger history of the Galaxy, representing streams of infalling matter or dynamical interaction with massive objects in the Galaxy.   

\acknowledgments

{We thank W.-C. Jao for useful discussion providing us with example spectra classified in \cite{jao08} system. We also thank the referee for his/her thorough comments that made this paper significantly better. 

A.S.S acknowledges the NASA Living with a Star Postdoctoral Fellowship. A.A.W acknowledges funding from NSF grants AST-1109273 and AST-1255568. A.A.W. also acknowledges the support of the Research Corporation for Science Advancement's Cottrell Scholarship. J.J.B. acknowledges the financial support of NSF grant AST 1151462. 

Funding for the SDSS and SDSS-II has been provided by the Alfred P. Sloan Foundation, the Participating Institutions, the National Science Foundation, the U.S. Department of Energy, the National Aeronautics and Space Administration, the Japanese Monbukagakusho, the Max Planck Society, and the Higher Education Funding Council for England. The SDSS Web Site is http://www.sdss.org/.

The SDSS is managed by the Astrophysical Research Consortium for the Participating Institutions. The Participating Institutions are the American Museum of Natural History, Astrophysical Institute Potsdam, University of Basel, University of Cambridge, Case Western Reserve University, University of Chicago, Drexel University, Fermilab, the Institute for Advanced Study, the Japan Participation Group, Johns Hopkins University, the Joint Institute for Nuclear Astrophysics, the Kavli Institute for Particle Astrophysics and Cosmology, the Korean Scientist Group, the Chinese Academy of Sciences (LAMOST), Los Alamos National Laboratory, the Max-Planck-Institute for Astronomy (MPIA), the Max-Planck-Institute for Astrophysics (MPA), New Mexico State University, Ohio State University, University of Pittsburgh, University of Portsmouth, Princeton University, the United States Naval Observatory, and the University of Washington.

\clearpage

\begin{sidewaystable}[h!]\tiny
\caption{Observed parameters of the { SDSS subdwarfs}.\label{tbl-1}}
\hspace{-2cm}\begin{tabular}{cccccccccccccccccc}
\tableline\tableline
ID & Plate & MJD & Fiber & RA & Dec & $u$ & $g$ & $r$ & $i$ & $z$ & $A_u$ & $A_g$ & $A_r$ & $A_i$ & $A_z$ & pmRA & pmDec \\
 &  &  &  & [deg] & [deg] & [mag] & [mag] & [mag] & [mag] & [mag] & [mag] & [mag] & [mag] & [mag] & [mag] & [mas\,yr$^{-1}$] & [mas\,yr$^{-1}$] \\
\tableline
SDSS012238.6$-$101651.7& 661 & 52163 & 187 & 20.66098 & -10.28101 & 26.12 & 20.99 & 19.19 & 17.92 & 17.16 & 0.21 & 0.15 & 0.11 & 0.08 & 0.06 & -27 & -14 \\
SDSS041007.3$-$041242.6 & 465 & 51910 & 587 & 62.53043 & -4.21183 & 23.01 & 20.91 & 19.14 & 18.22 & 17.72 & 0.36 & 0.26 & 0.19 & 0.14 & 0.10 & 54 & -195 \\
SDSS073406.5+363731.9 & 431 & 51877 & 90 & 113.52729 & 36.62552 & 22.92 & 20.94 & 19.28 & 18.31 & 17.77 & 0.27 & 0.20 & 0.15 & 0.11 & 0.08 & 2 & -2  \\
SDSS123433.7+663950.5 & 494 & 51915 & 51 & 188.64022 & 66.66404 & 22.44 & 19.57 & 17.84 & 17.01 & 16.52 & 0.07 & 0.05 & 0.04 & 0.03 & 0.02 & -505 & -215 \\
SDSS124636.2+665006.8 & 495 & 51988 & 262 & 191.65093 & 66.83521 & 24.97 & 21.62 & 19.89 & 18.85 & 18.32 & 0.12 & 0.09 & 0.06 & 0.05 & 0.03 & 15 & 12 \\
SDSS125635.9$-$001944.9 & 293 & 51994 & 183 & 194.14963 & -0.32915 & 24.69 & 21.42 & 19.51 & 18.24 & 17.52 & 0.13 & 0.10 & 0.07 & 0.05 & 0.04 & -9999 & -9999 \\
SDSS130331.6$-$030708.7 & 339 & 51692 & 225 & 195.88181 & -3.11909 & 23.70 & 20.66 & 19.21 & 18.27 & 17.70 & 0.15 & 0.11 & 0.08 & 0.06 & 0.04 & -22 & 22 \\
SDSS132147.1+014804.0& 526 & 52312 & 104 & 200.44627 & 1.80110 & 22.32 & 19.98 & 18.52 & 17.78 & 17.23 & 0.16 & 0.12 & 0.08 & 0.06 & 0.05 & -8 & 12\\
SDSS155823.0+533229.2 & 621 & 52055 & 338 & 239.59565 & 53.54143 & 25.02 & 22.75 & 20.62 & 19.54 & 18.94 & 0.07 & 0.05 & 0.04 & 0.03 & 0.02 & -81 & -171 \\
\tableline
\end{tabular}
\end{sidewaystable} 

\clearpage

\begin{sidewaystable}[h!]\tiny
\begin{center}
\caption{Derived parameters of the { SDSS subdwarfs}.\label{tbl-2}}
\hspace{-2cm}\begin{tabular}{ccccccccccccccccccccc}
\tableline\tableline
ID & CaH2 & CaH3 & TiO5 & RV & $\zeta$ & Sp & Metal. & $M_r$ & $M_i$ & $M_z$ & $D$ & $z_{gal}$ & $U$ & $V$ & $W$ & $V_{\mathrm{t}}$ & $V_{tot}$ & Act. & EW H$\alpha$ \\
& & & & [km\,s$^{-1}$] & & & class & [mag] & [mag] & [mag] & [pc] & [pc] & [km\,s$^{-1}$] & [km\,s$^{-1}$] & [km\,s$^{-1}$] & [km\,s$^{-1}$] & [km\,s$^{-1}$] & & [\AA] \\
\tableline
SDSS012238.6$-$101651.7& 0.40 & 0.62 & 0.49 & 12 & 0.780 & 4 & sdM & 12.47 & 11.08 & 10.51 & 209 & -184 & 37 & 19 & -11 & 31 & 33 & -9999 & 1.33\\
SDSS041007.3$-$041242.6& 0.41 & 0.58 & 0.86 & 126 & 0.205 & 4 & esdM & 11.34 & 10.42 & 10.06 & 332 & -186 & 51 & -296 & -139 & 320 & 344 & 0 & -0.7 \\
SDSS073406.5+363731.9& 0.54 & 0.75 & 0.66 & 50 & 0.742 & 2 & sdM & 11.01 & 10.05 & 9.62 & 423 & 187 & -33 & 5 & 30 & 6 & 50 & 0 & 0.4 \\
SDSS123433.7+663950.5& 0.47 & 0.65 & 0.89 & 104 & 0.192 & 3 & usdM & 11.41 & 10.49 & 10.14 & 190 & 162 & -346 & -300 & 180 & 495 & 506 & 0 & 0.5 \\
SDSS124636.2+665006.8& 0.50 & 0.69 & 0.59 & -29 & 0.763 & 2 & sdM & 11.76 & 10.70 & 10.27 & 412 & 332 & 37 & 28 & -29 & 38 & 48 & 0 & -0.1\\
SDSS125635.9$-$001944.9& 0.31 & 0.41 & 0.59 & -9999 & 0.483 & 6 & esdM & 13.07 & 11.66 & 11.08 & 188 & 182 & -9999 & -9999 & -9999 & -9999 & -9999 & -9999 & -9999 \\
SDSS130331.6$-$030708.7& 0.53 & 0.74 & 0.64 & 5 & 0.772 & 2 & sdM & 10.07 & 9.15 & 8.68 & 650 & 575 & -75 & 21 & 49 & 96 & 96 & 0 & -0.6 \\
SDSS132147.1+014804.0& 0.67 & 0.83 & 0.79 & 79 & 0.647 & 1 & sdM & 9.66 & 8.88 & 8.43 & 572 & 527 & 5 & 2 & 95 & 39 & 88 & 0 & -0.1  \\
SDSS155823.0+533229.2 & 0.42 & 0.63 & 0.69 & -175 & 0.499 & 3 & esdM & 14.15 & 12.88 & 12.45 & 194 & 156 & 113 & -208 & -38 & 174 & 247 & -9999 & 1.9  \\
\tableline
\end{tabular}
\end{center}
\end{sidewaystable} 

\clearpage

\begin{table}[h!]
\begin{center}
\caption{Spectral lines used for RV determinations.\label{tbl-3}}
\begin{tabular}{cccc}
\tableline\tableline
Line & $\lambda_{vac}$ [\AA] & Line & $\lambda_{vac}$ [\AA]\\
\tableline
KI  & 7667.0089 & NaI & 8185.5054\\
KI  & 7701.0825 & NaI & 8197.0766\\
RbI & 7949.7890 & TiI & 8437.2600\\
RbI & 7802.4140 &&\\
\tableline
\end{tabular}
\begin{tablenotes}
      \small
      \item The values are in vacuum wavelengths and are adopted from NIST.
    \end{tablenotes}
\end{center}
\end{table}

\clearpage

\begin{table}[h!]
\begin{center}
\caption{Spectral features used in spectral typing. \label{tbl-4}}
\begin{tabular}{ccc}
\tableline\tableline
Band & Continuum 1 [\AA] & Feature [\AA]\\
\tableline
TiO5  & 7042-7046 & 7126-7135\\
CaH2  & 7042-7046 & 6814-6846\\
CaH3  & 7042-7046 & 6960-6990\\
\tableline
\end{tabular}
\begin{tablenotes}
      \small
      \item Spectral features and partial continua for the determination of the strength of the CaH2, CaH3, and TiO5 spectral features. Values are in vacuum wavelengths.
    \end{tablenotes}
\end{center}
\end{table}

\clearpage

\begin{table}[h!]
\begin{center}
\caption{Mean colors per metallicity and spectral subclass,\label{tbl-5}}
\begin{tabular}{cccccc}
\tableline\tableline
Metalicity & Spectral type & Mean $g-r$ ($\sigma$) & Mean $r-z$ ($\sigma$)\\
\tableline
dM   & All & 1.52 (0.47) & 1.39 (1.77)\\
\tableline 
sdM  & 0-1 & 1.46 (0.23) & 1.21 (0.21)\\
     & 2-3 & 1.55 (0.22) & 1.66 (0.25)\\
     & 4-5 & 1.61 (0.28) & 2.09 (0.40)\\
     & 6-7 & 1.74 (0.23) & 2.70 (0.67)\\
\tableline
esdM & 0-1 & 1.48 (0.18) & 1.13 (0.21)\\
     & 2-3 & 1.67 (0.22) & 1.38 (0.22)\\
     & 4-5 & 1.79 (0.13) & 1.56 (0.22)\\
     & 6-7 & 1.99 (0.16) & 1.84 (0.016)\\
\tableline
usdM & 0-1 & 1.55 (0.23) & 1.24 (0.38)\\
     & 2-3 & 1.72 (0.18) & 1.28 (0.17)\\
     & 4-5 & 1.86 (0.10) & 1.42 (0.14)\\
\tableline
\end{tabular}
\end{center}
\end{table}

\clearpage

\begin{table}[h!]
\begin{center}
\caption{Mean RPM per metallicity and spectral subclass,\label{tbl-hr}}
\begin{tabular}{cccccc}
\tableline\tableline
Metalicity & Spectral type & Mean RPM & $\sigma_{\mathrm{RPM}}$\\
\tableline
dM   & All & 12.7 & 5.6 \\
\tableline
sdM  & All & 16.6 & 2.6 \\ 
     & 0-1 & 15.9 & 2.6 \\
     & 2-3 & 16.5 & 2.3 \\
     & 4-5 & 18.3 & 2.6 \\
     & 6-7 & 19.7 & 2.5 \\     
\tableline
esdM & All & 18.0 & 2.7 \\
     & 0-1 & 17.0 & 2.4 \\
     & 2-3 & 19.1 & 2.3 \\
     & 4-5 & 20.6 & 1.4 \\
     & 6-7 & 21.7 & 1.3 \\
\tableline
usdM & All & 18.3 & 2.7 \\
     & 0-1 & 17.6 & 2.6 \\
     & 2-3 & 19.2 & 2.3 \\
     & 4-5 & 20.9 & 0.9 \\
\tableline
\end{tabular}
\end{center}
\end{table}

\clearpage

\begin{table}[h!]\footnotesize
\begin{center}
\caption{Velocities and velocity dispersions for the three metallicity classes of subdwarfs and the three Galactic components. \label{tbl-6}}
\begin{tabular}{ccccc}
\tableline\tableline
Class & Mean velocity$^{\mathrm{a}}$ & Velocity dispersion$^{\mathrm{a}}$ & Mean velocity$^{\mathrm{b}}$ & Velocity dispersion$^{\mathrm{b}}$ \\
& [km\,s$^{-1}$] & [km\,s$^{-1}$] & [km\,s$^{-1}$] & [km\,s$^{-1}$] \\
\tableline
     & $\langle U\rangle$=8   & $\sigma_U$=101 & $\langle U\rangle$= 6 & $\sigma_U$=68 \\
sdM  & $\langle V\rangle$=-54 & $\sigma_V$=115 & $\langle V\rangle$=-53 & $\sigma_V$=92 \\
     & $\langle W\rangle$=4   & $\sigma_W$=82  & $\langle W\rangle$=3 & $\sigma_W$=47 \\
\tableline     
     & $\langle U\rangle$=8    & $\sigma_U$=101 & $\langle U\rangle$=16 & $\sigma_U$=140 \\
esdM & $\langle V\rangle$=-155 & $\sigma_V$=132 & $\langle V\rangle$=-180 & $\sigma_V$=114 \\
     & $\langle W\rangle$=2    & $\sigma_W$=118 & $\langle W\rangle$=-1 & $\sigma_W$=71 \\
\tableline
     & $\langle U\rangle$=6   & $\sigma_U$=161  & $\langle U\rangle$=20 & $\sigma_U$=131 \\
usdM & $\langle V\rangle$=-171 & $\sigma_V$=152 & $\langle V\rangle$=-144 & $\sigma_V$=117 \\
     & $\langle W\rangle$=10   & $\sigma_W$=111 & $\langle W\rangle$=-18 & $\sigma_W$=77 \\ 
\tableline
     & $\langle U\rangle$=-9   & $\sigma_U$=32  & & \\
Thin disk$^{\mathrm{c}}$  & $\langle V\rangle$=-20 & $\sigma_V$=23 & & \\
     & $\langle W\rangle$=-7   & $\sigma_W$=18  & & \\
     \tableline
     & $\langle U\rangle$=-10   & $\sigma_U$=49  \\
Thick disk$^{\mathrm{c}}$   & $\langle V\rangle$=-32 & $\sigma_V$=38 & & \\
     & $\langle W\rangle$=-7   & $\sigma_W$=40  & & \\
\tableline
     & $\langle U\rangle$=0   & $\sigma_U$=135  & & \\
Halo$^{\mathrm{d}}$  & $\langle V\rangle$=-173 & $\sigma_V$=105 & & \\
     & $\langle W\rangle$=0   & $\sigma_W$=90  & & \\   
\tableline
\end{tabular}
\begin{tablenotes}
      \small
      \item $^{\mathrm{a}}$ { The values for the mean velocities and their dispersions for the three metalliciy classes are taken from this study}
      \item $^{\mathrm{b}}$ { The values for the mean velocities and their dispersions for the three metalliciy classes are taken from \cite{bochanski12} Table\,5 and 6 from the color category that most closely corresponds to the mean color for the different mettalicity classes.}  
      \item $^{\mathrm{c}}$ The values for the mean velocities and their dispersions for { the thin and thick disk Galactic components} are taken from \cite{fuchs09}. 
      \item $^{\mathrm{d}}$ { The values for the mean velocities and their dispersions for the halo Galactic component are taken from \cite{binney98}}. 
    \end{tablenotes}
\end{center}
\end{table}

\clearpage

\begin{sidewaystable}[h!]\tiny
\caption{Kinematic parameters for the fast ($>$ 525\,km\,s$^{-1}$) subdwarfs in the sample.\label{tbl-7}}
\hspace{-2cm}\begin{tabular}{ccccccccccccc}
\tableline\tableline
ID & Plate & MJD & Fiber & pmRA & pmDec & RV & $D$ & $U$ & $V$ & $W$ & $V_{\mathrm{t}}$& $V_{tot}$\\
 &  &  &  & [marcsec\,yr$^{-1}$] & [marcsec\,yr$^{-1}$] & [km\,s$^{-1}$] & [pc] & [km\,s$^{-1}$] & [km\,s$^{-1}$] & [km\,s$^{-1}$] & [km\,s$^{-1}$] & [km\,s$^{-1}$] \\
\tableline
SDSS080301.1+354848.4 & 757 & 51997 & 3 & 20 & -489 & 105 & 303 & -87 & -688 & -64 & 704 & 711 \\
SDSS105717.3+462102.2 & 1436 & 51913 & 379 & 75 & -527 & -296 & 199 & 321 & -517 & -84 & 545 & 621 \\
SDSS113018.7$-$030506.4 & 327 & 51694 & 305 & -174 & -206 & 47 & 706 & -163 & -727 & -479 & 902 & 903 \\
SDSS120840.1+192834.4 & 2918 & 51788 & 148 & -105 & -114 & 231 & 652 & -118 & -490 & 130 & 481 & 533 \\
SDSS121441.2+414924.8 & 1450 & 51994 & 151 & -517 & -447 & -110 & 228 & -238 & -689 & -58 & 740 & 748 \\
SDSS122842.6$-$023247.4 & 334 & 51692 & 494 & 171 & -279 & 116 & 447 & 626 & -289 & -153 & 694 & 704 \\
SDSS125135.7+581841.7 & 2461 & 52339 & 127 & -151 & -60 & 12 & 1162 & -536 & -672 & 190 & 896 & 896 \\
SDSS141730.9+180014.3 & 2759 & 51900 & 426 & 85 & -9 & -65 & 1458 & 432 & 342 & -257 &  594 & 598 \\
SDSS142259.7+143754.5 & 2746 & 52347 & 144 & -150 & -336 & -91 &  328 & 134 & -540 & -117 & 573 & 580 \\
SDSS143526.0+383305.6 & 1349 & 52314 & 86 & -257 & -197 & -125 & 358 & -41 & -539 & 111 & 550 & 564 \\
SDSS144846.9+614802.3 & 609 & 51997 & 123 & 154 & -973 & -104 & 144 & 608 & -257 & 193 & 674 & 682 \\
SDSS150211.7+353152.9 & 1384 & 51994 & 518 & -174 & -246 & -356 & 289 & -73 & -640 & -69 & 559 & 663 \\
SDSS151420.9+351335.8 & 1353 & 51692 & 123 & -278 & 181 & -193 & 424 & -633 & -222 & 117 & 668 & 695 \\
SDSS151534.5+312819.6 & 1649 & 52000 & 481 & -372 & -143 & -141 & 278 & -152 & -488 & 140 & 525 & 544 \\
\tableline
\end{tabular}
\end{sidewaystable} 

\clearpage

\begin{figure}
\epsscale{.90}
\plotone{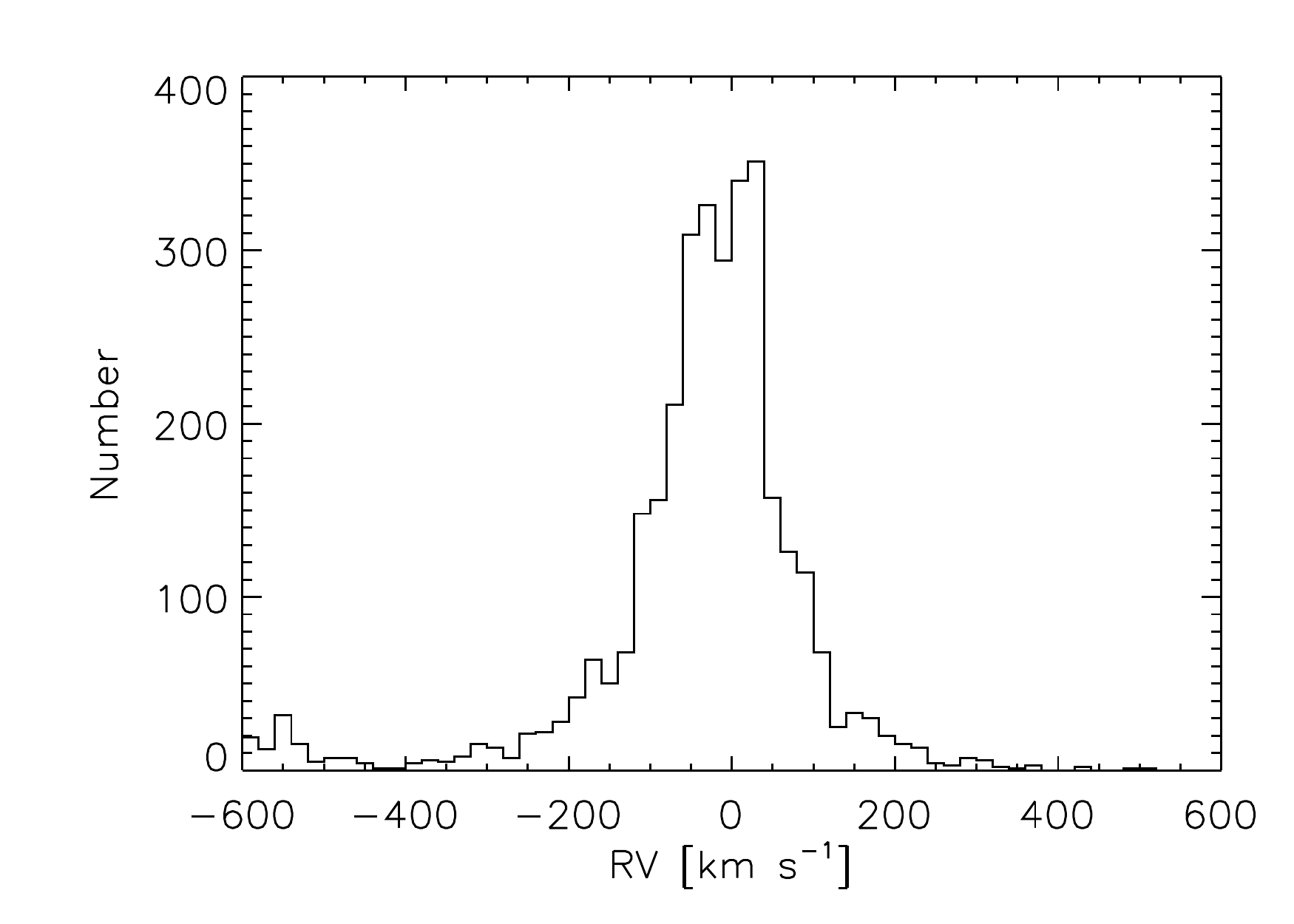}
\caption{Distribution of all radial velocities in the sample. The distribution peaks at zero but is broad with some stars having radial velocity of several hundred km\,s$^{-1}$. \label{rvhist}}
\end{figure}

\clearpage

\begin{figure}
\epsscale{.90}
\plotone{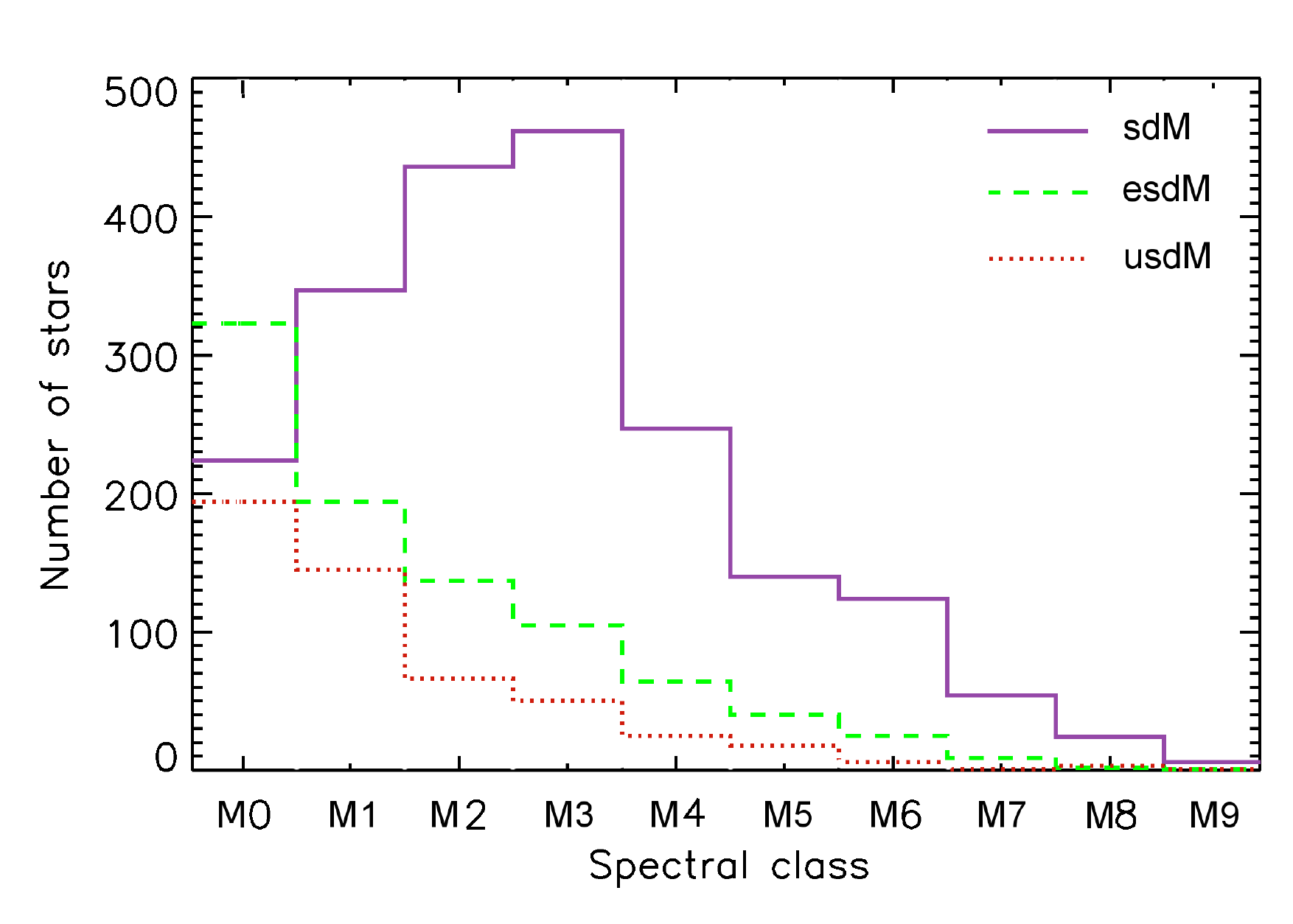}
\caption{Distributions of spectral subclass for all metallicity classes - solid purple line for sdMs, green dashed for esdMs, and red dotted for usdMs. Most sdMs are found in spectral classes 2 and 3, while the most esdMs and usdMs are in spectral class 1. All stars in the sample are taken into account. \label{spstat}}
\end{figure}

\clearpage

\begin{figure}
\epsscale{.90}
\plotone{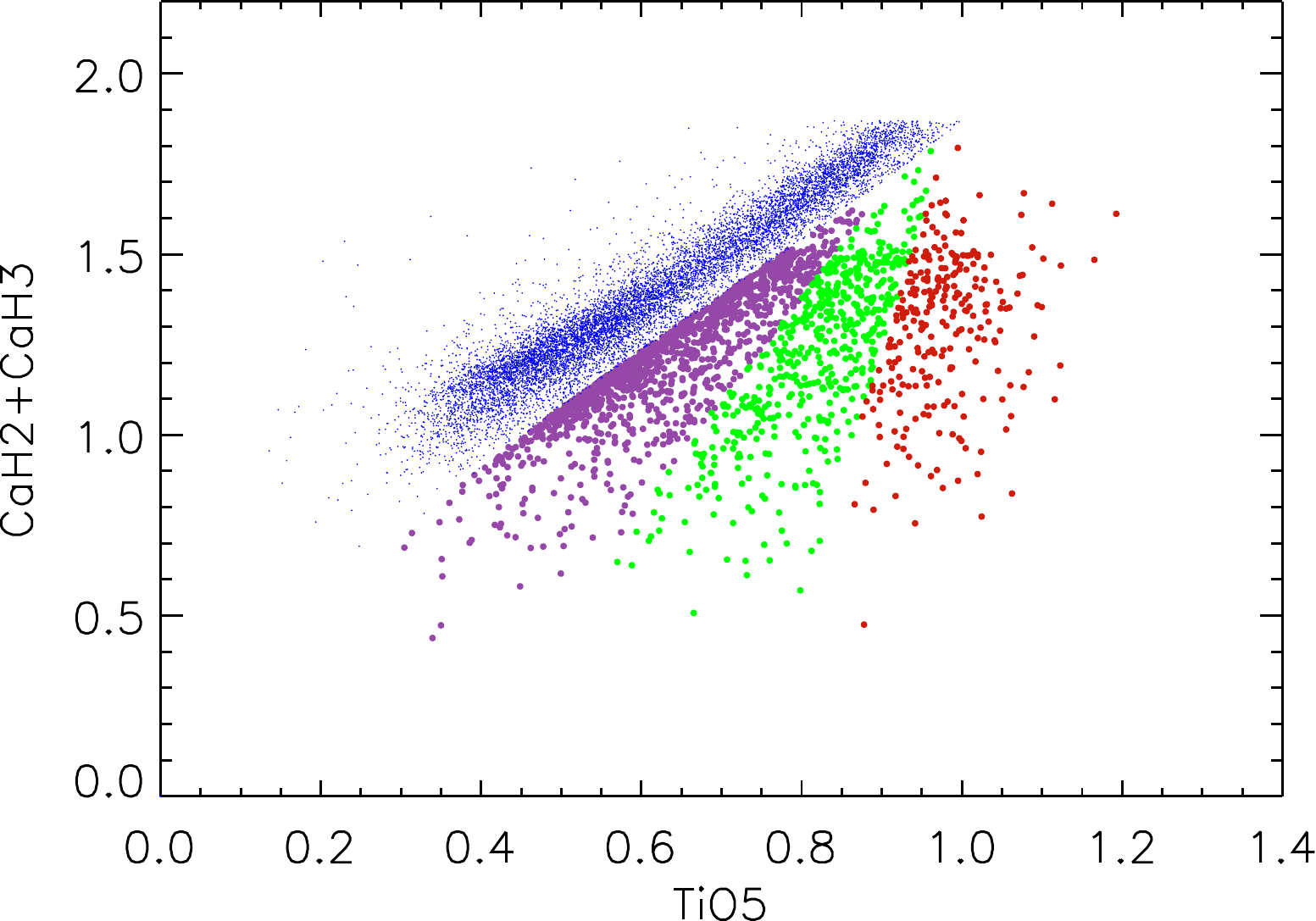}
\caption{Distribution of dMs, sdMs , esdMs, and usdMs in CaH-TiO space. dMs are blue, sdMs are purple, esdMs - green, and usdMs -- red, corresponding to the colors in Figure\,1. All stars in the sample are plotted. The field M dwarfs are M dwarfs from \cite{west11} with proper motions larger than 30\,mas\,yr$^{-1}$. \label{ind}}
\end{figure}

\clearpage

\begin{figure}
\epsscale{.80}
\plotone{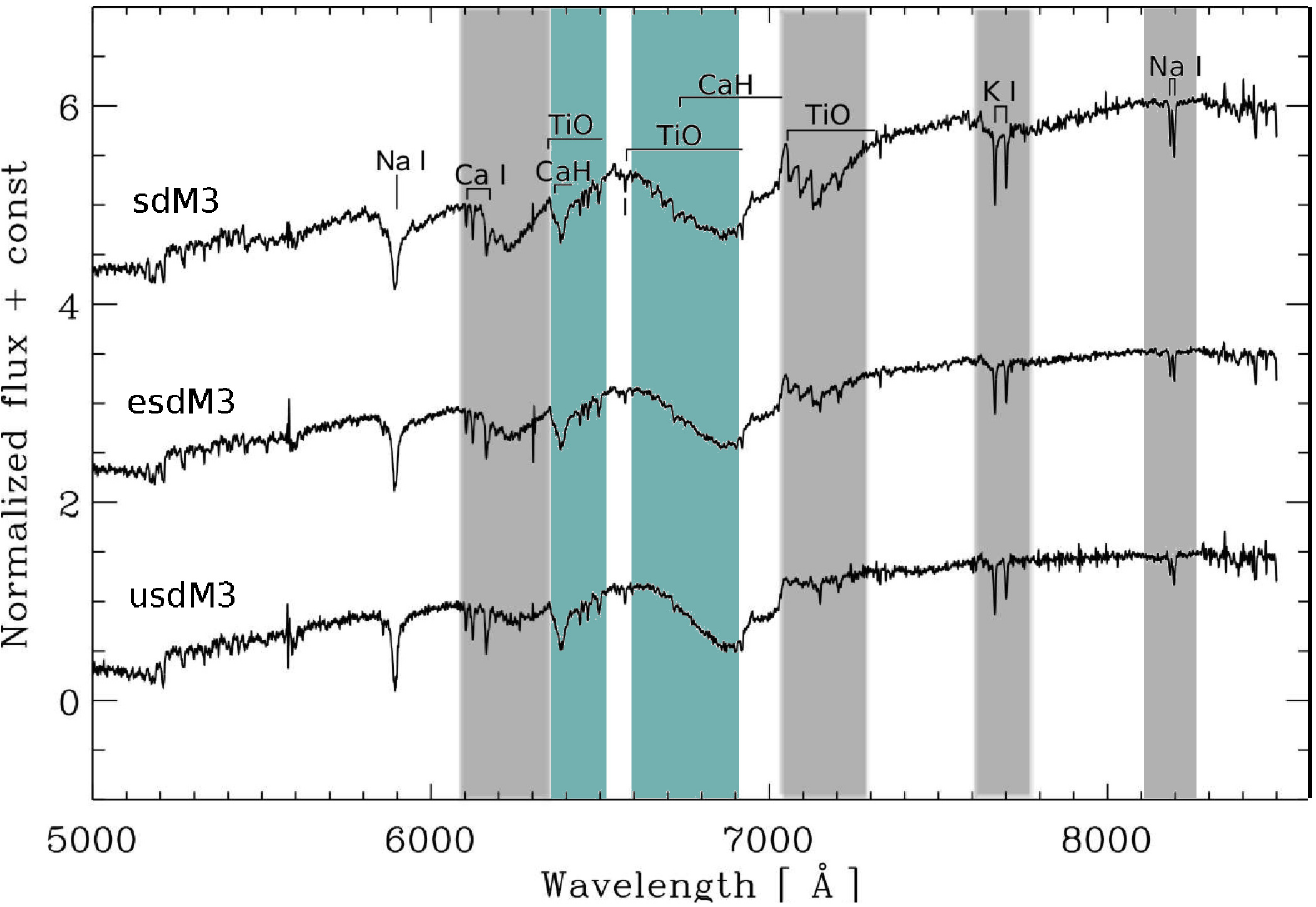}
\caption{Sample template spectra for sdM3 (top), esdM3 (middle), and usdM3 (bottom) that show how various spectral features change as a function of metallicty. The spectral regions where the metallicity effects are most obvious are shown as shaded gray regions. The cyan colored regions mark the bands where temperature effects may play a significant role. Most of the regions are likely affected by both metalliciy and temparature. Prominent spectral features are marked on the top spectrum.\label{3temp}}
\end{figure}

\clearpage

\begin{figure}
\epsscale{.80}
\plotone{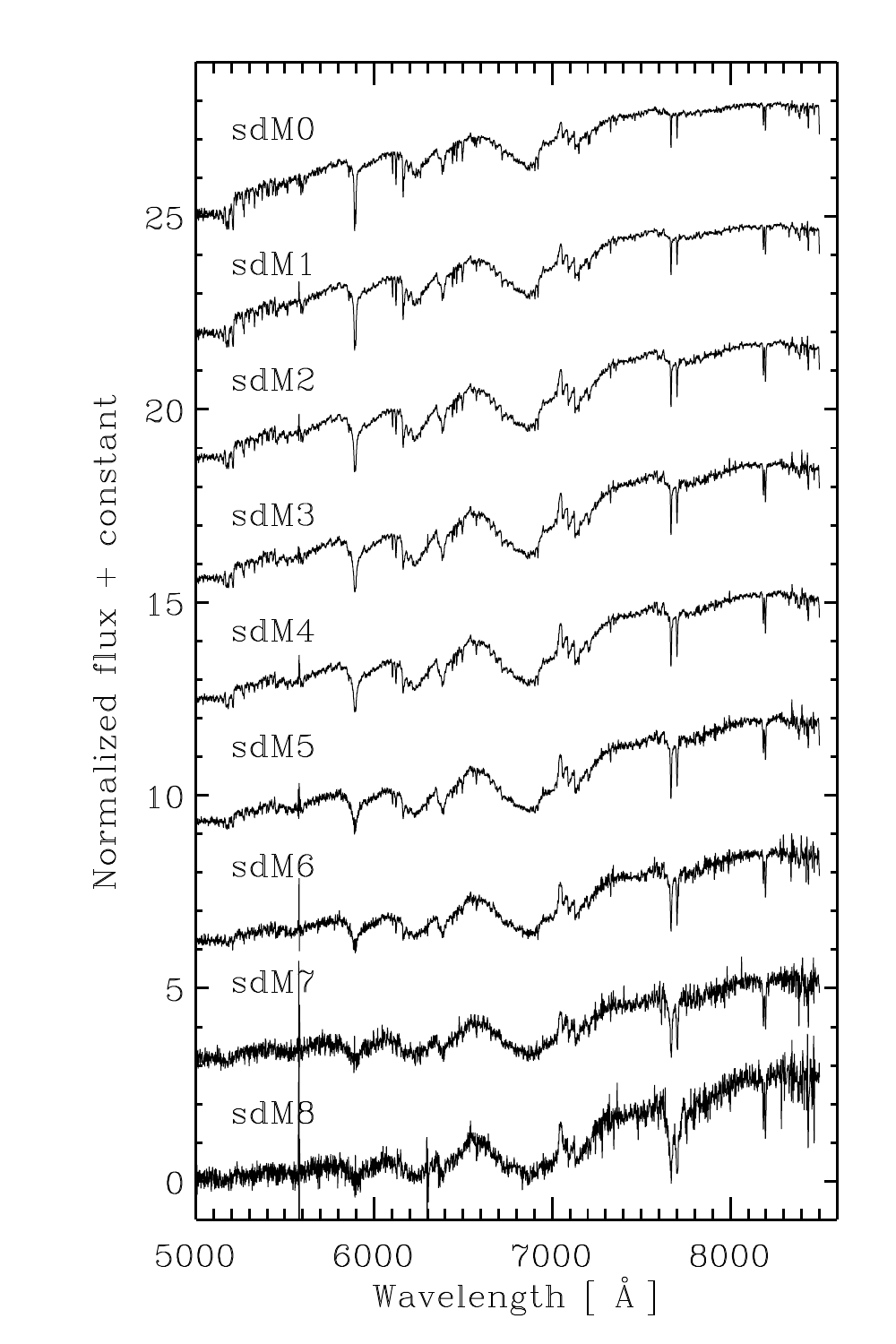}
\caption{Template spectra for sdMs for spectral classes sdM0 to sdM8 in ascending order. All spectra have been normalized, multiplied by a constant and a constant has been added to represent them on the same figure.\label{temp1}}
\end{figure}

\clearpage

\begin{figure}
\epsscale{.80}
\plotone{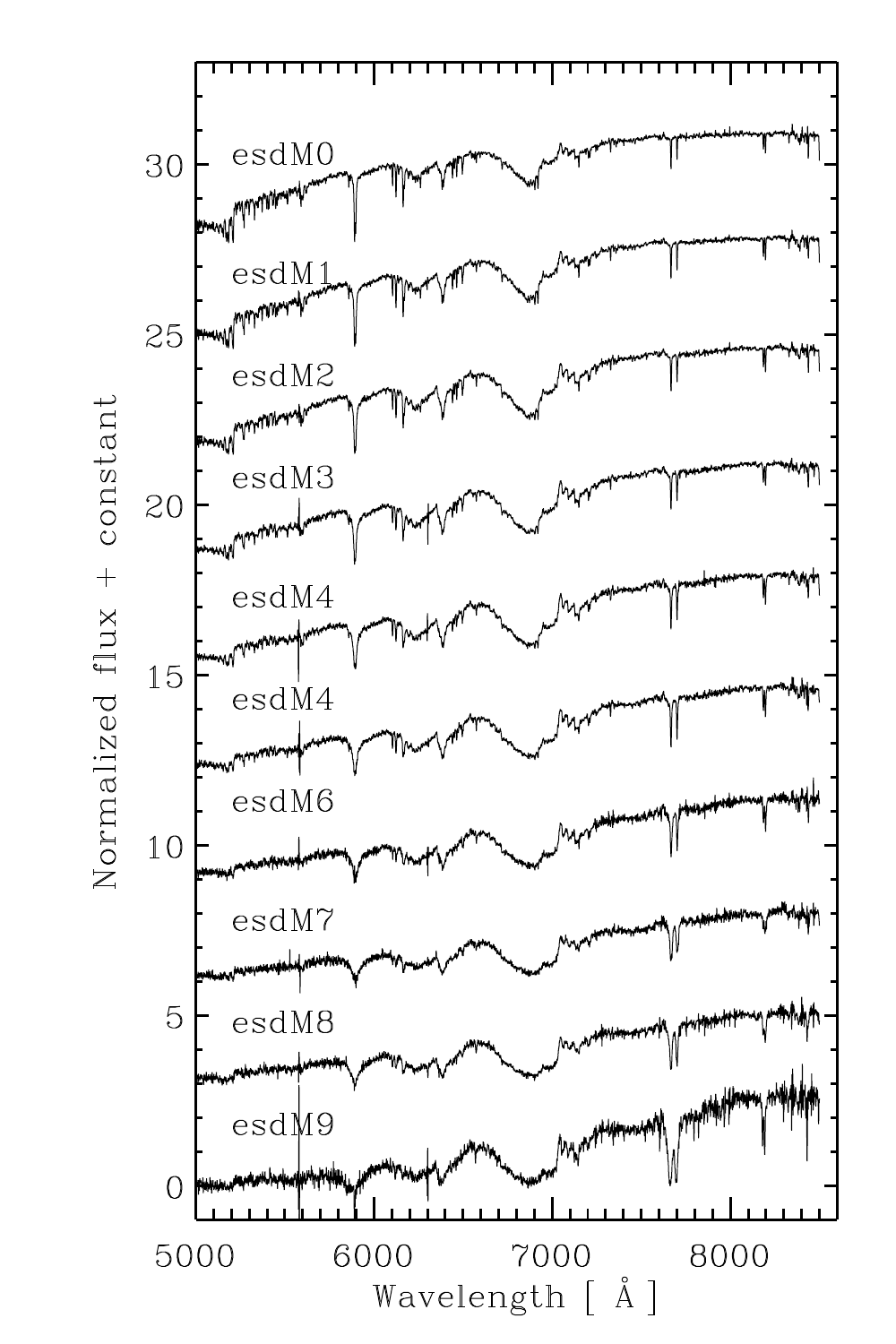}
\caption{Template spectra for esdMs for spectral classes sdM0 to sdM9 in ascending order.\label{temp2}}
\end{figure}

\clearpage

\begin{figure}
\epsscale{.80}
\plotone{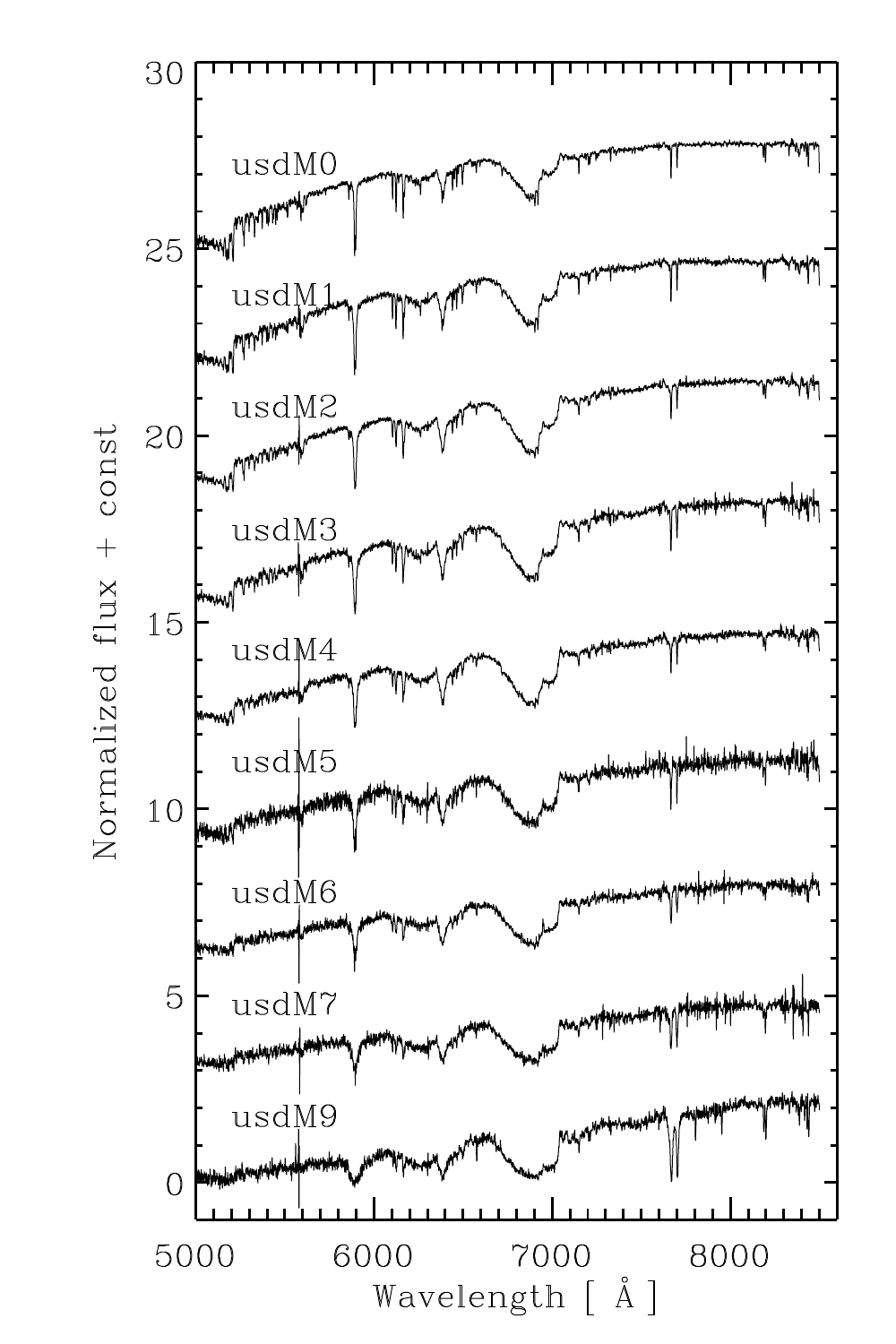}
\caption{Same as Fig.\,2, but for usdMs. The template for the usdM8 is not shown due to the lack of enough stars to make a good template.\label{temp3}}
\end{figure}

\clearpage

\begin{figure}
\epsscale{.80}
\plotone{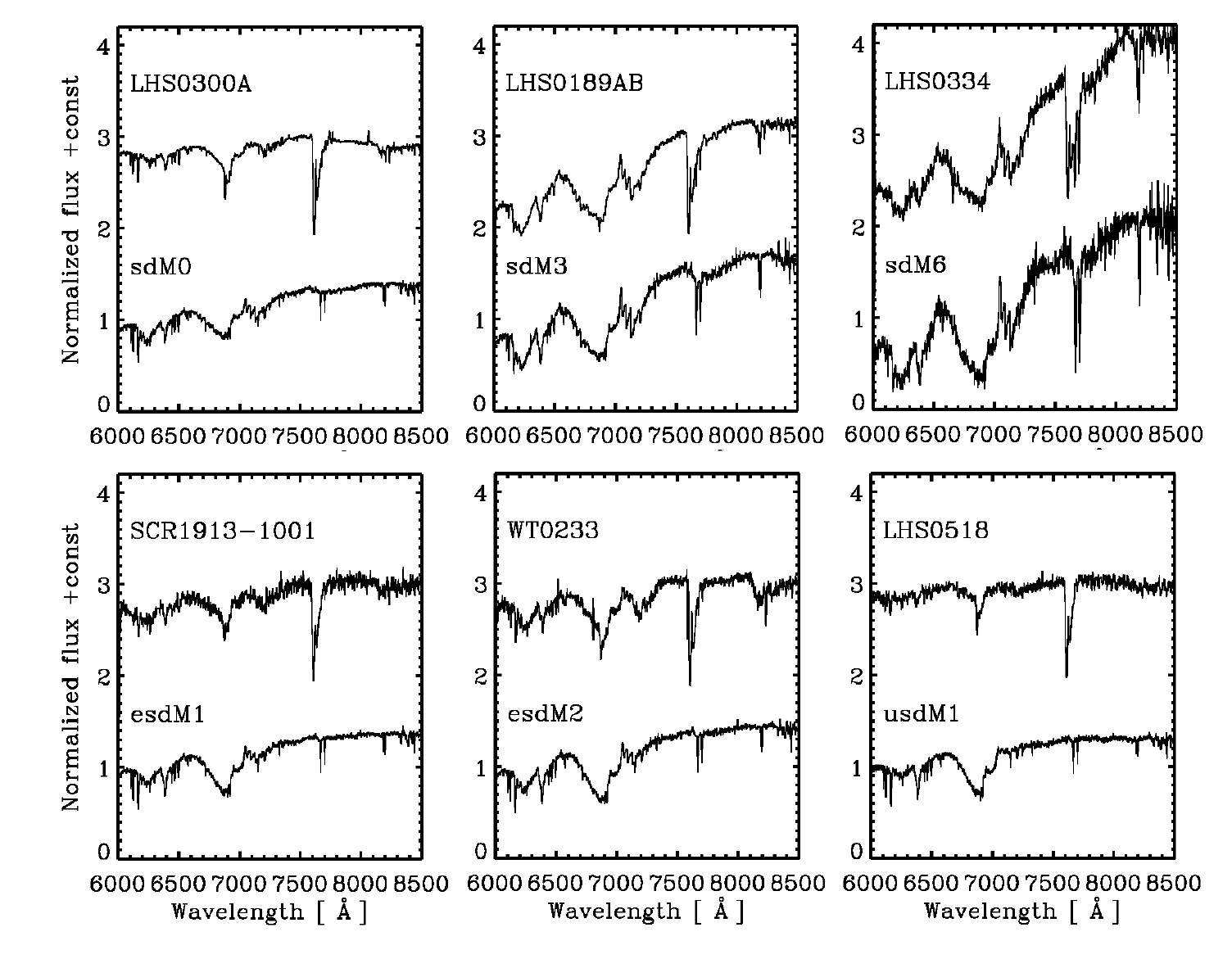}
\caption{Comparison between templates for different metallicity and spectral subtype from this work (bottom spectra) and spectra classified using the \cite{jao08} system (top spectra).These example spectra are described in \cite{jao08}. The \cite{jao08} spectra are taken on the basis of their determined spectral subtype { (which matches that of the template spectra)} and metallicity indicator since they are not classified as sdM, esdM and usdM in the Jao system -- { we take one/two levels of metallicity to} correspond to sdM, three/four -- esdM, and five/six -- usdM.\label{tempjao}}
\end{figure}

\clearpage

\begin{figure}
\epsscale{.90}
\plotone{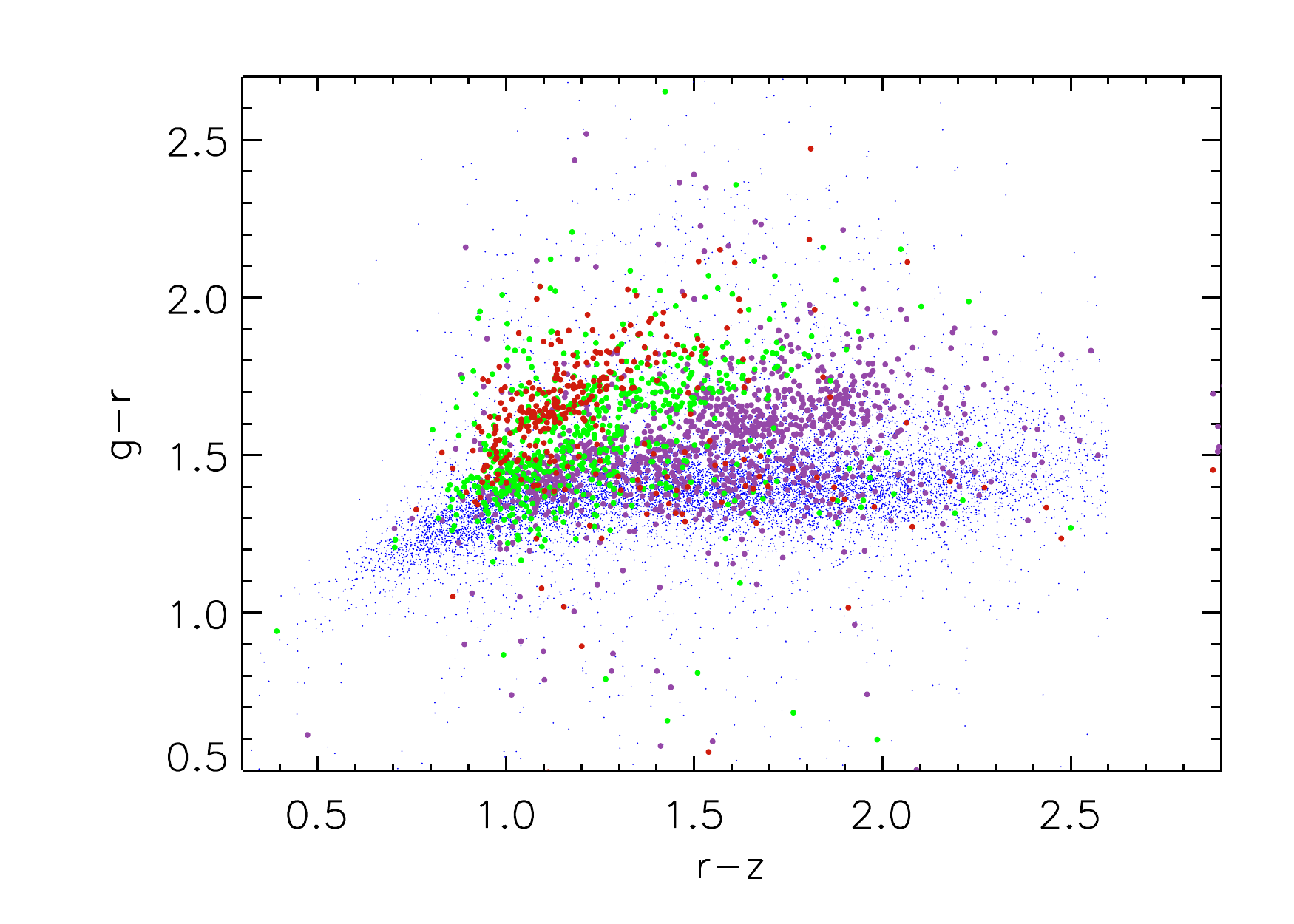}
\caption{Color-color diagram of subdwarfs (purple for sdM, green for esdM, and red for usdM) and field M dwarfs with high proper motions { from \cite{west11}} as blue dots. { The colors have been corrected for extinction.} The different metallicity classes separate in color with respect to the field dMs and among each other although there is some overlap. The field M dwarfs are the same as in Fig.\,\ref{ind}. Only the stars with proper motions are plotted.\label{ccd}}
\end{figure}

\clearpage

\begin{figure}
\epsscale{.90}
\plotone{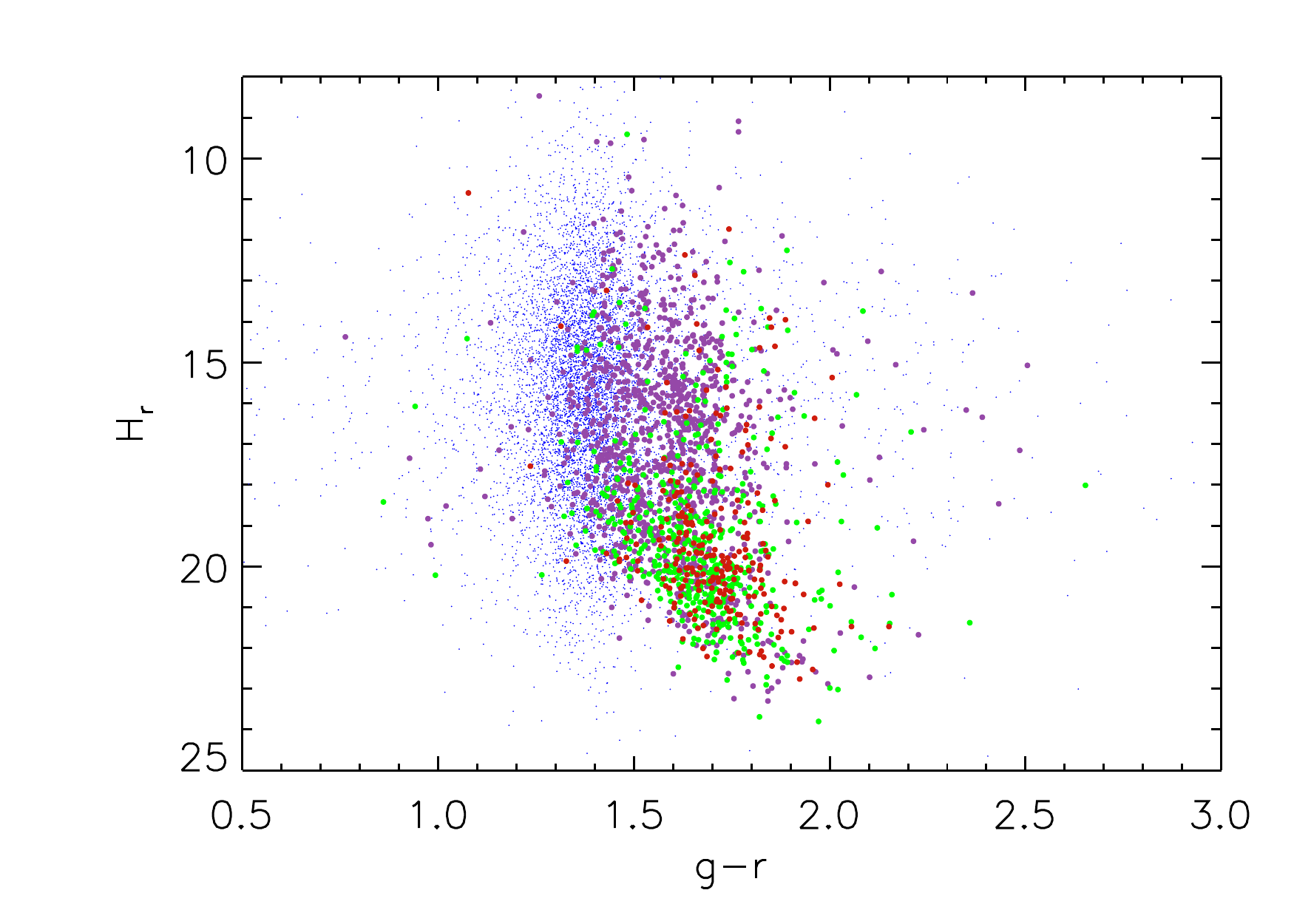}
\caption{RPM versus $g-r$ for all subdwarfs { with proper motions in the sample (2368 stars). { The colors have been corrected for extincton.} We use eq.\,(2) to compute the reduced proper motion in this figure.} The color coding is the same as in Figure\,\ref{ccd}. There is a good separation of the subdwarfs and field M dwarfs on the diagram. The RPMs of dMs and sdMs in this diagram are similar, linking the sdMs with the Galactic disk, while esdMs and usdMs have preferentially larger RPM values consistent with them being members of the stellar halo. The field M dwarfs are the same as in Fig.\,\ref{ind}. \label{ccrpm}}
\end{figure} 

\clearpage

\begin{figure}
\epsscale{.60}
\plotone{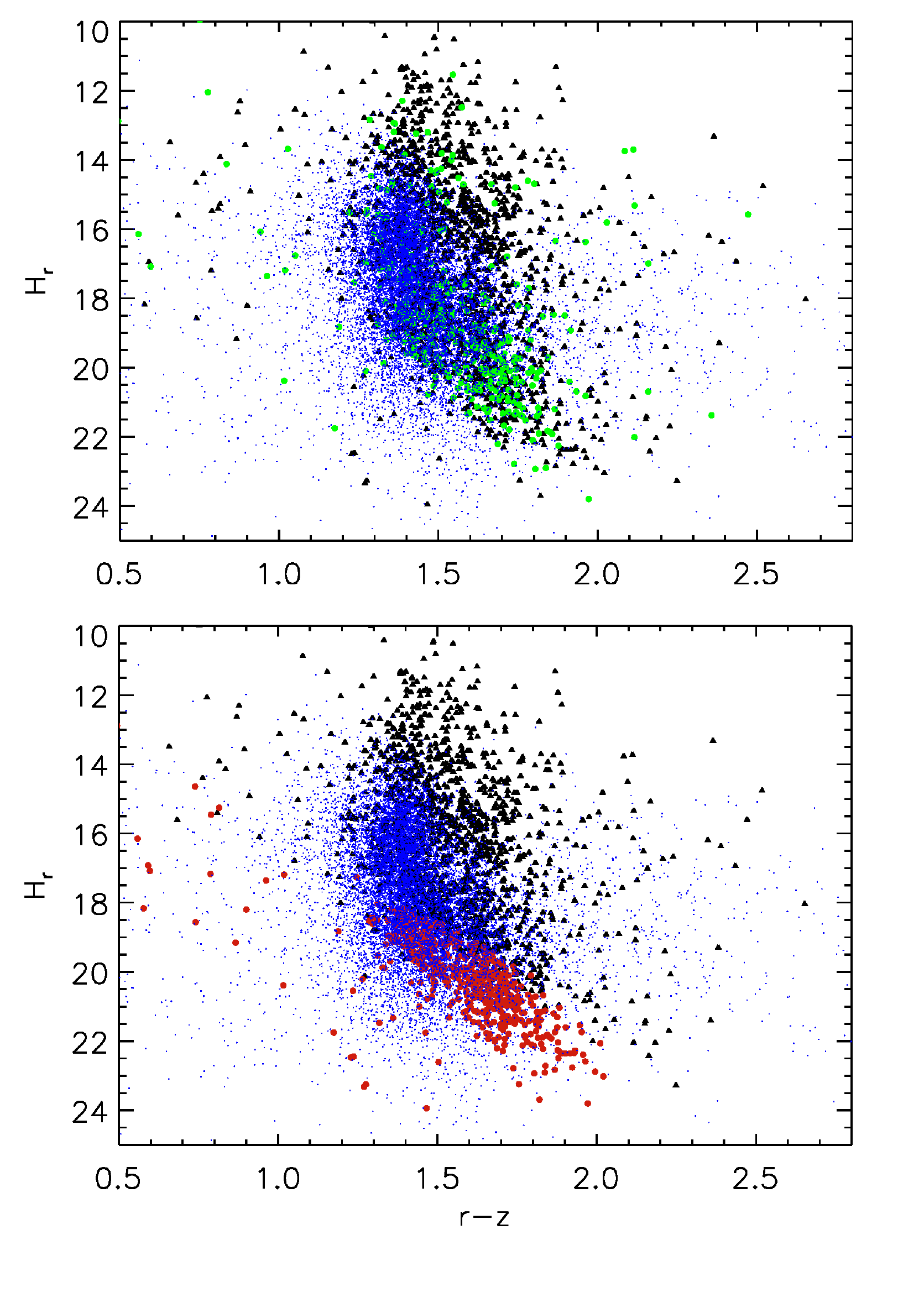}
\caption{Top: Reduced proper motion diagram as in Figure\,\ref{ccrpm}, but with the stars with radial velocity larger than 150\,km\,s$^{-1}$ represented with green circles. { The blue dots are field M dwarfs from \cite{west11}. Only stars with proper motions are plotted (2368 stars). The reduced proper motion in both panels is computed using eq.\,(2).} The rest of the stars are black triangles. Stars with large radial velocities also exhibit larger proper motions, suggesting they are moving quickly through the Galaxy and are likely members of the stellar halo. Bottom: The same reduced proper motion diagram, but now the stars with transverse velocities larger than 200\,km\,s$^{-1}$ are represented with red circles. The other stars are black triangles. Again, stars with large tangential velocities, and hence proper motions, occupy the region with large RPM values. Many stars belong to both subsets. \label{rvrpm}}
\end{figure}

\clearpage

\begin{figure}
\epsscale{.90}
\plotone{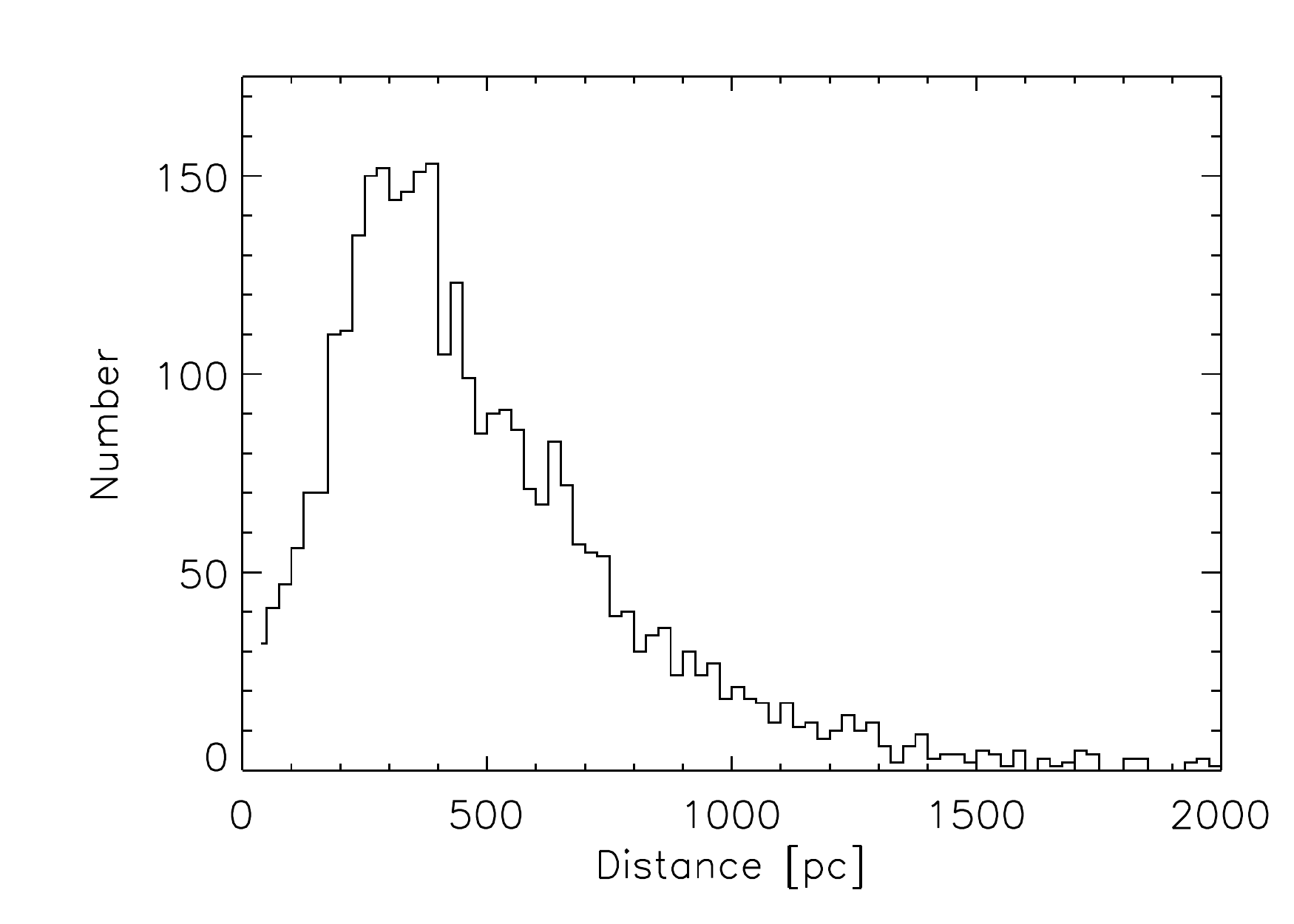}
\caption{Distribution of distances to all the subdwarfs in the sample, which have been determined based on { the absolute magntude in the $r$-band} from statistical parallax method of \cite{bochanski12}. The majority of the stars are within 200-400\,pc, with an extended tail reaching 2000\,pc. The typical uncertainty in the absolute magnitude is 0.4\,magnitudes resulting in mean distance uncertainty of 20\%. \label{dist}}
\end{figure}

\clearpage

\begin{figure}
\epsscale{.90}
\plotone{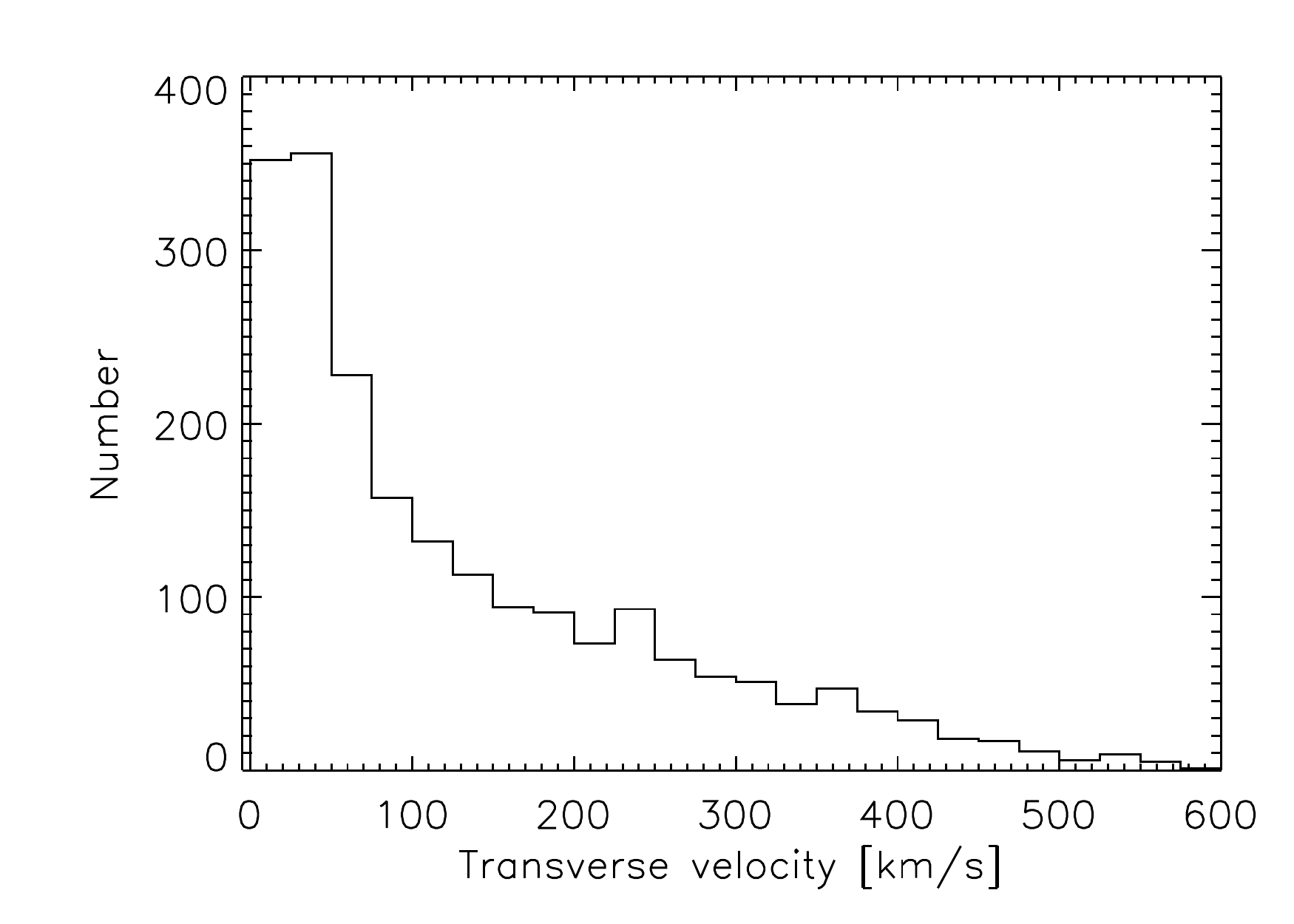}
\caption{Distribution of transverse velocities of all subwarfs in the sample, determined based on the proper motions and the distances shown above. { Only stars with proper motions are plotted.} The histogram peaks at low velocities, but there is a significant tail reaching 800\,km\,s$^{-1}$, although the last bins may be due to low S/N data which shows us with inrealistically high velocities. \label{vthist}}
\end{figure}

\clearpage

\begin{figure}
\epsscale{.90}
\plotone{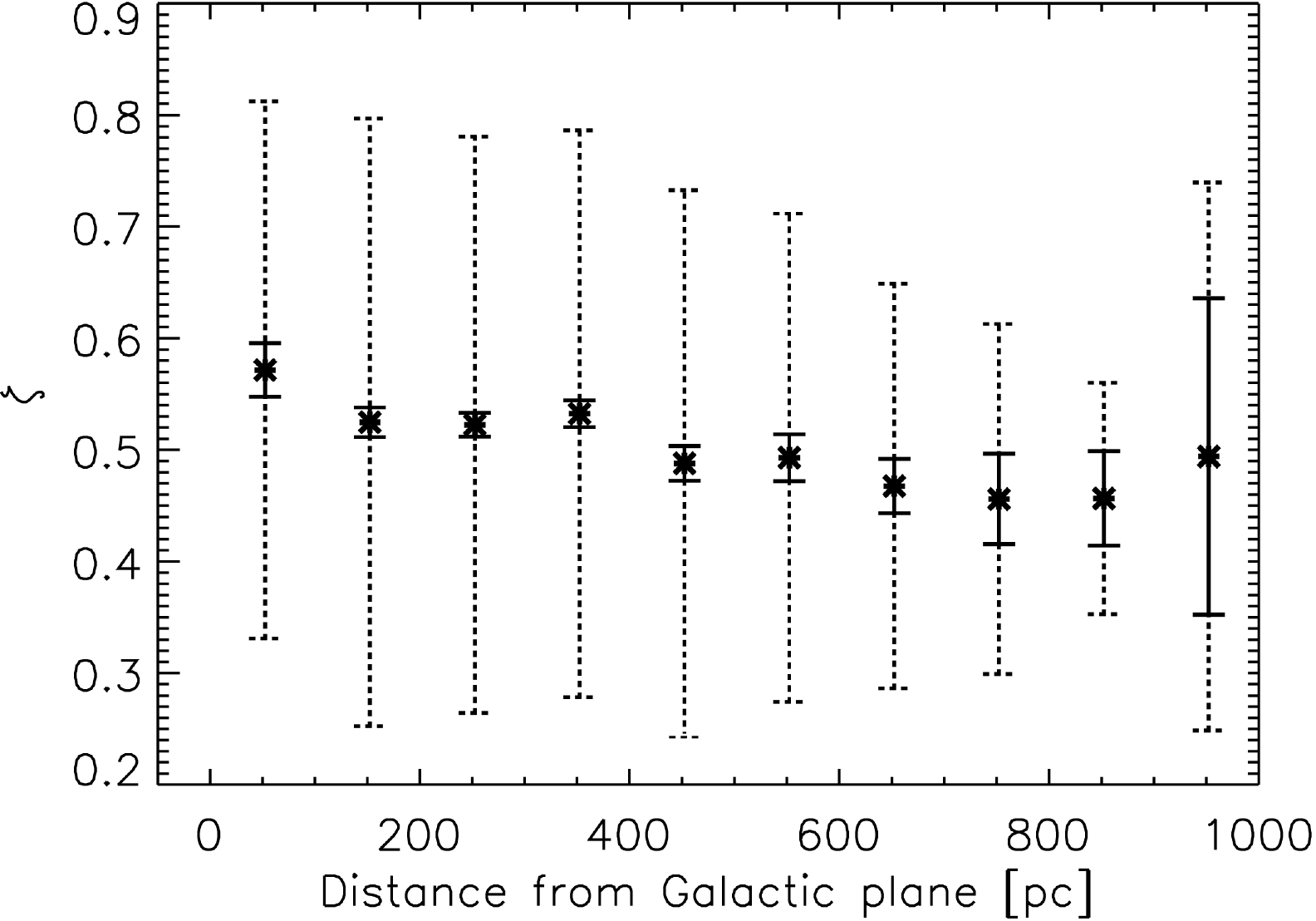}
\caption{Mean values of the metallicity proxy, $\zeta$ in bins of distance from the galactic plane. The error in the mean is shown as the solid error bars, and the standard deviation is the dashed error bar. There is a slight decrease of $\zeta$ with distance from the plane. The spread in $\zeta$ is significant, which points to a mix of different metalicity classes at all distances from the Galactic plane, with generally more spread at smaller distances, pointing to more mixing closer to the plane. \label{zeta}}
\end{figure}

\clearpage

\begin{figure}
\epsscale{.90}
\plotone{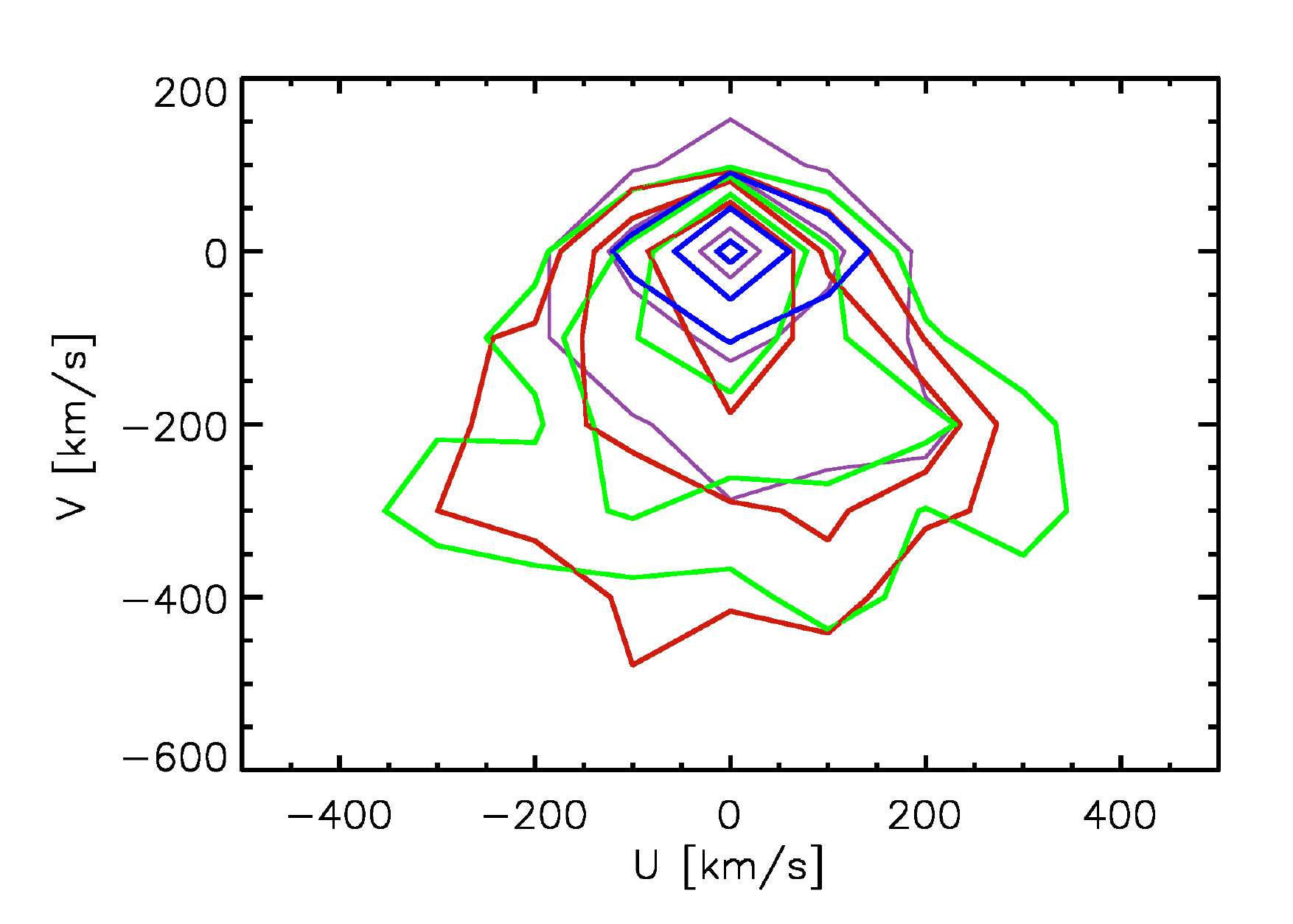}
\caption{Galactic velocities $V$ versus $U$ velocities for all subdwarfs and field M dwarfs with high proper motions. The contours are at { 40, 68 and 95\%. The} M dwarf distribution is given in blue contours, sdMs -- purple, esdMs -- green, usdMs -- red contours. The dispersion in $V$ is largest and the whole distribution is offset to large velocities for the usdMs and esdMs, while the contours for sdMs are centered at zero. \label{vu}}
\end{figure}
\clearpage

\begin{figure}
\epsscale{.60}
\plotone{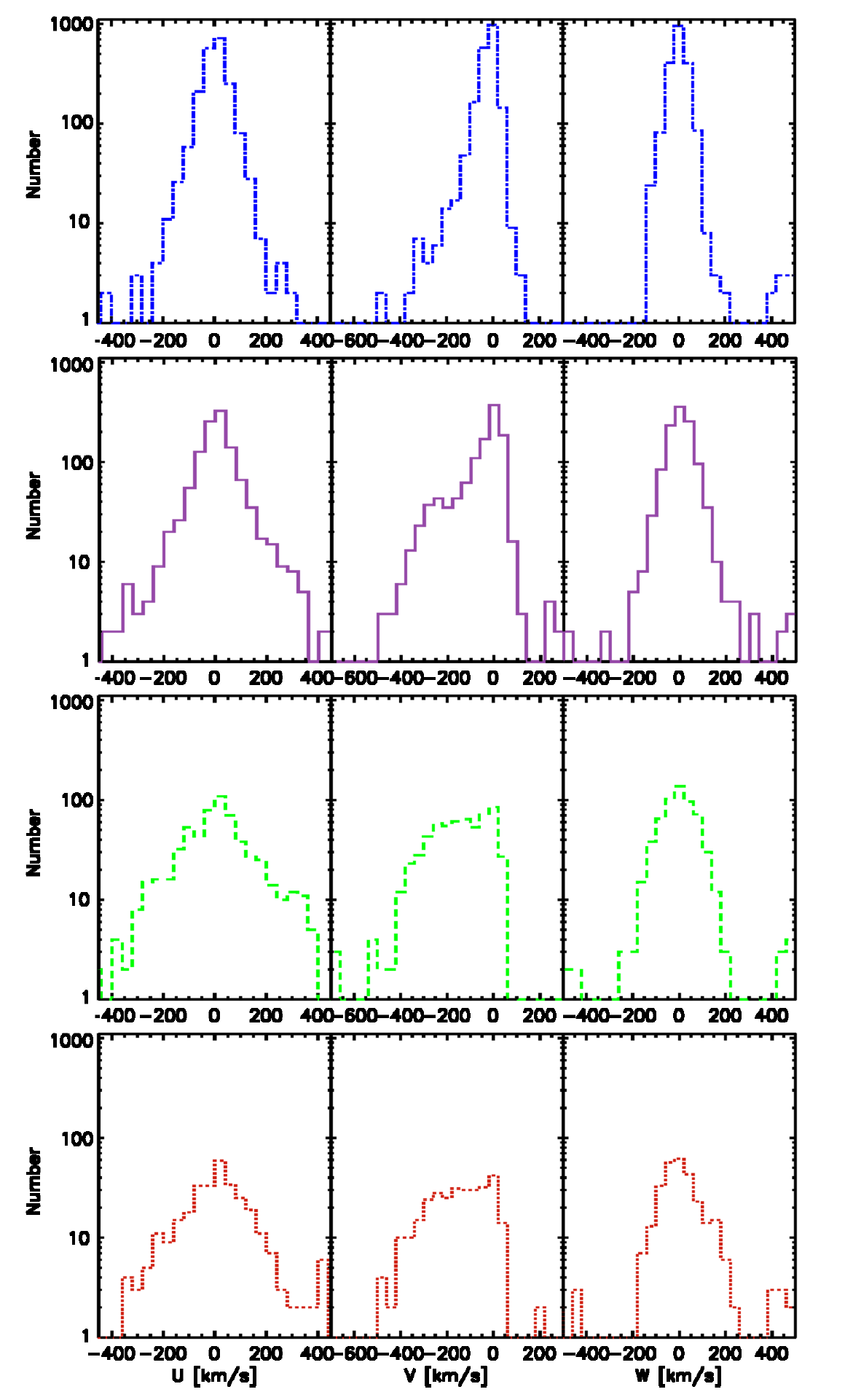}
\caption{Galactic velocity histograms on a logarithmic scale for all subdwarfs and field M dwarfs. The red and green (esdMs and usdMs respectively) histograms of the $V$ velocity component peak at around $-170$\,km\,s$^{-1}$ while the sdMs (purple) and field M dwarfs (blue) peak around 0\,km\,s$^{-1}$. \label{uvwhist}}
\end{figure}
\clearpage

\begin{figure}
\epsscale{.60}
\plotone{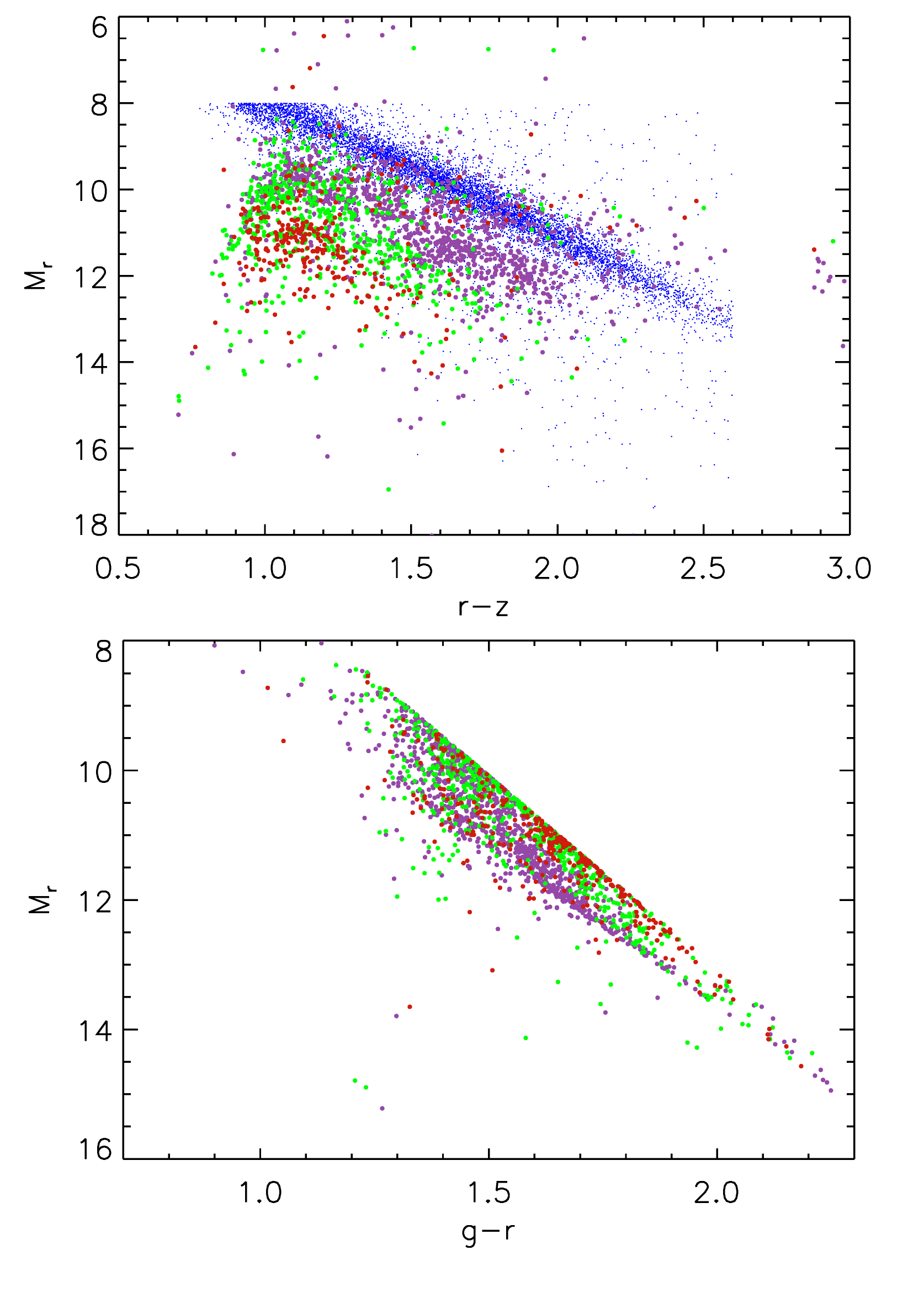}
\caption{Top: Color-absolute magnitude diagram of $M_R$ vs $r-z$ color for all subdwarfs (color coding is the same as in Figure\,9). { The colors have been corrected for extinction.} As expected, subdwarfs lie under (or to the left of) the main sequence field M dwarfs. The different metalicity classes separate well on this diagram, with the ordinary subdwarfs spanning a large range in absolute magnitude. Bottom: Color-magnitude diagram of $M_r$ vs. $g-r$ color. The distributions of subdwarfs is much tighter in absolute magnitude in this CMD with the opposite dependence on metallicity. This relation is much tighter than the top plot since the distribution in color in $g-r$ is also much tighter (see Figure\,\ref{ccd}.) There is a sharp edge to the subdwarfs since $M_r$ is a function of $g-r$ in a given range \citep{bochanski12}.\label{Mr-zr}}
\end{figure}

\clearpage

\begin{figure}
\epsscale{.90}
\plotone{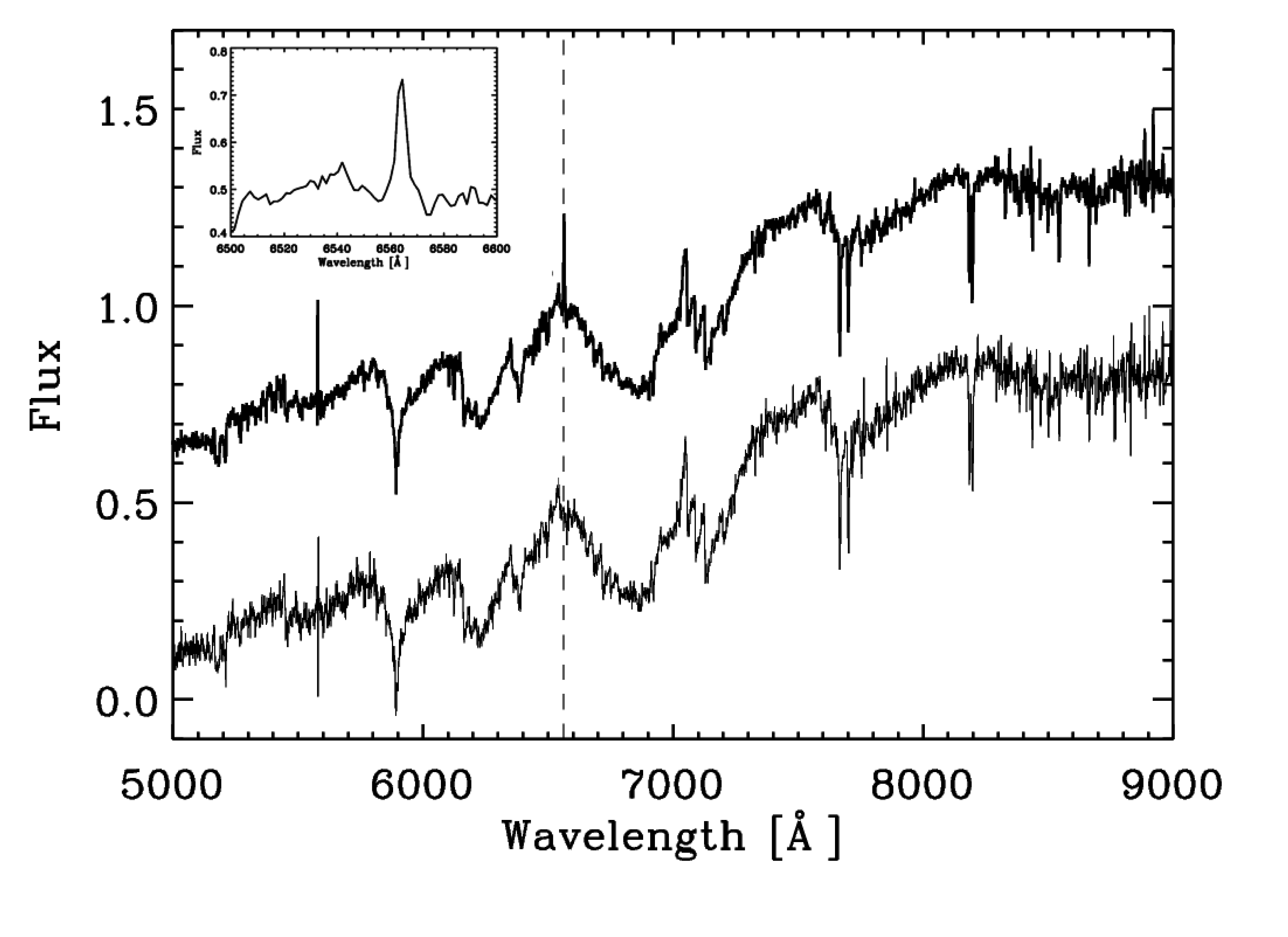}
\caption{Sample spectra of inactive (bottom) and an active (top) sdM3 stars. A vertical dashed line is plotted at the location of the H${\alpha}$ line. A zoom-in on the region around the H${\alpha}$ line for the active star is shown in the inset panel.\label{act}}
\end{figure}

\clearpage

\begin{figure}
\epsscale{.90}
\plotone{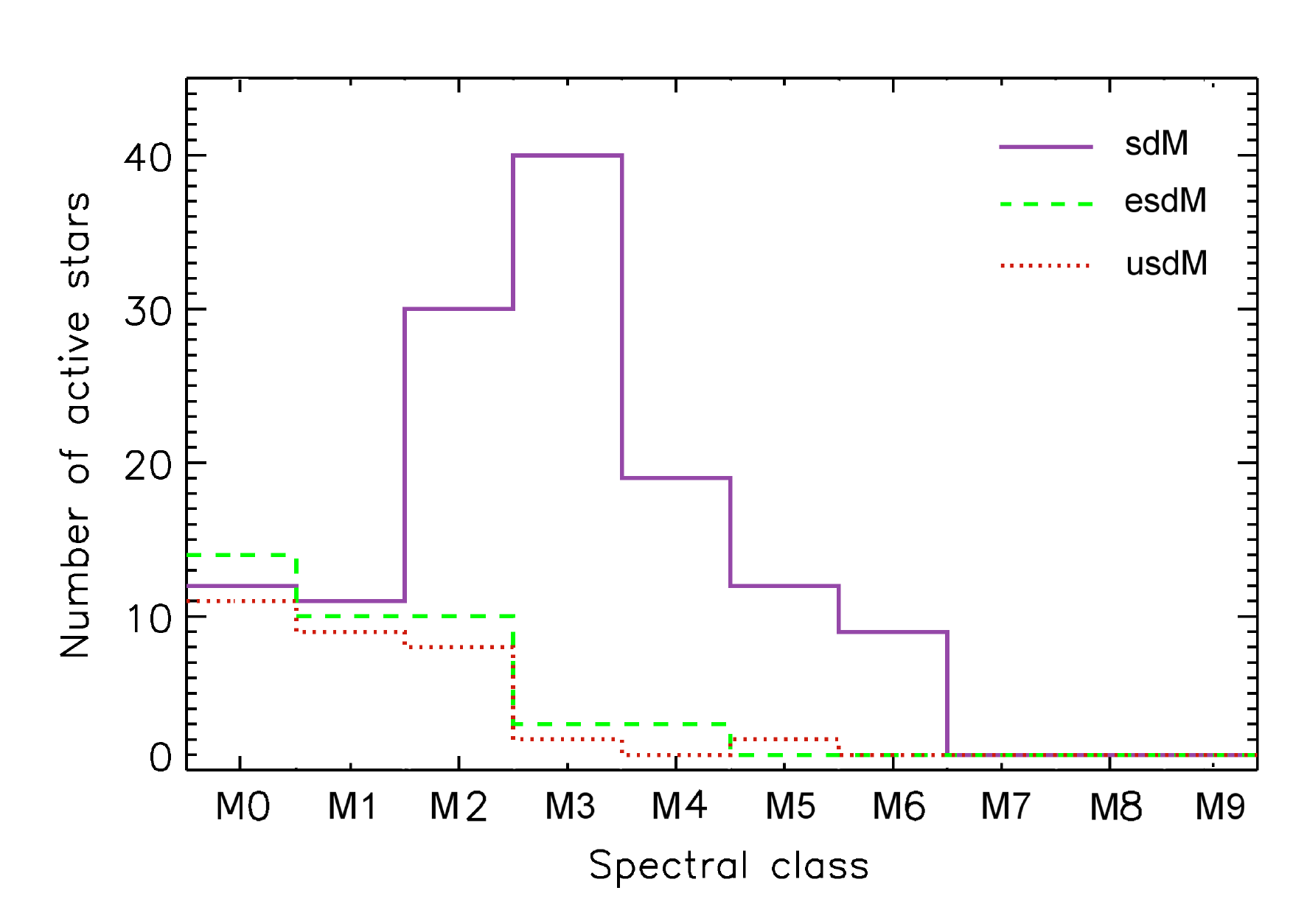}
\caption{Distribution of active stars in the different spectral classes for the three metallicity classses. The colors correspond to previous figures - purple for sdMs, green for esdMs, and red for usdMs. The most active sdMs are found in spectral class 3, the most active esdMs and usdMs are in spectral classes 2 and 1 respectively.\label{actsp}}
\end{figure}

\clearpage

\begin{figure}
\epsscale{.90}
\plotone{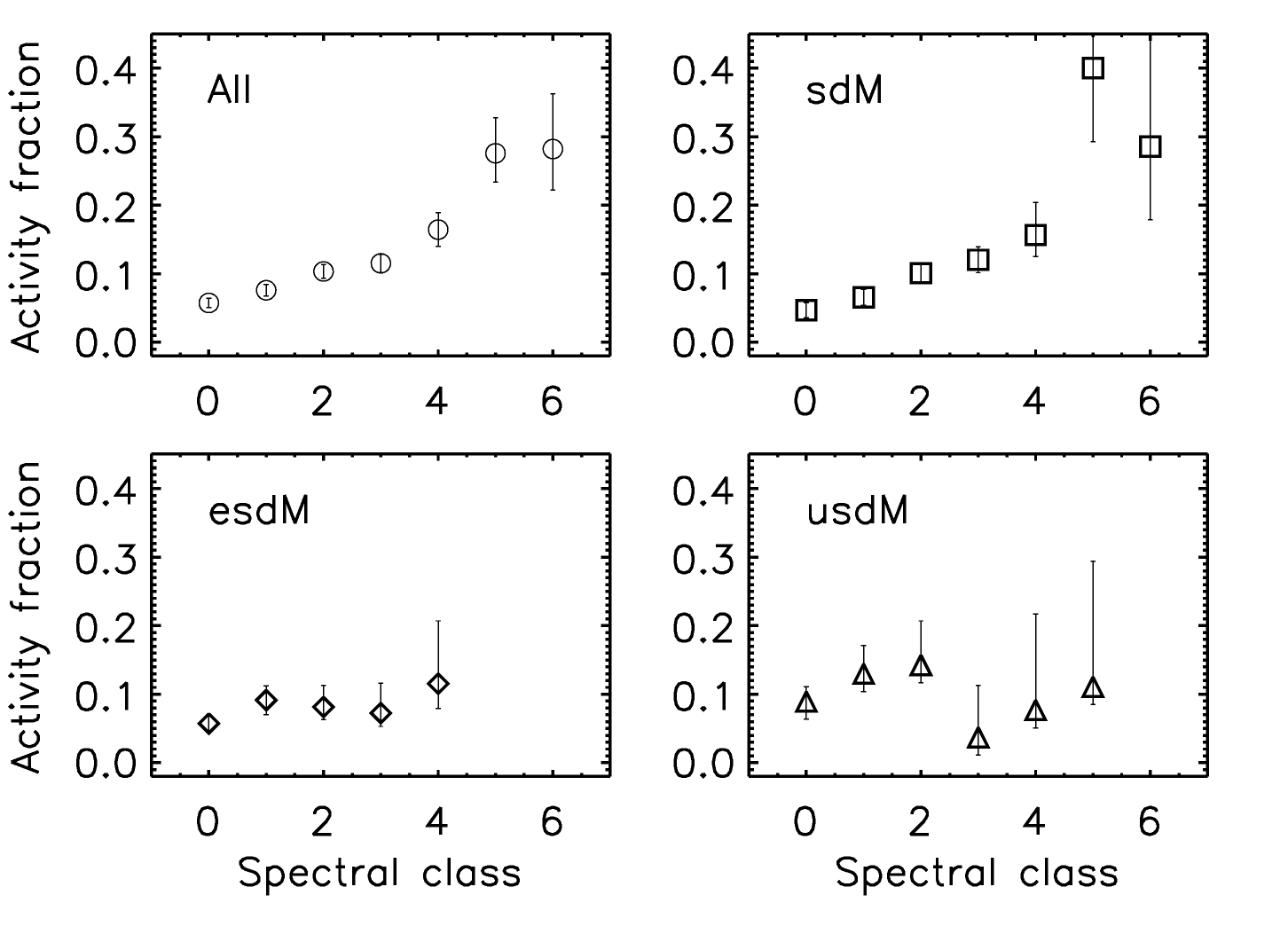}
\caption{Activity fractions in bins of spectral subtype for the three metallicity classes. { All definitely active and inactive stars (activity=1 and 0 respectively) from the sample are included.} The plot in the top left includes all sdMs, esdMs, and usdMs. The uncertainties, shown on the figure, are computed based on binomial statistics. We see a clear trend of dependence of activity on spectral class for the sdMs (similar to normal disk dwarfs), but not for the more metal poor subdwarfs. \label{actspbin}}
\end{figure}

\clearpage

\begin{figure}
\epsscale{.90}
\plotone{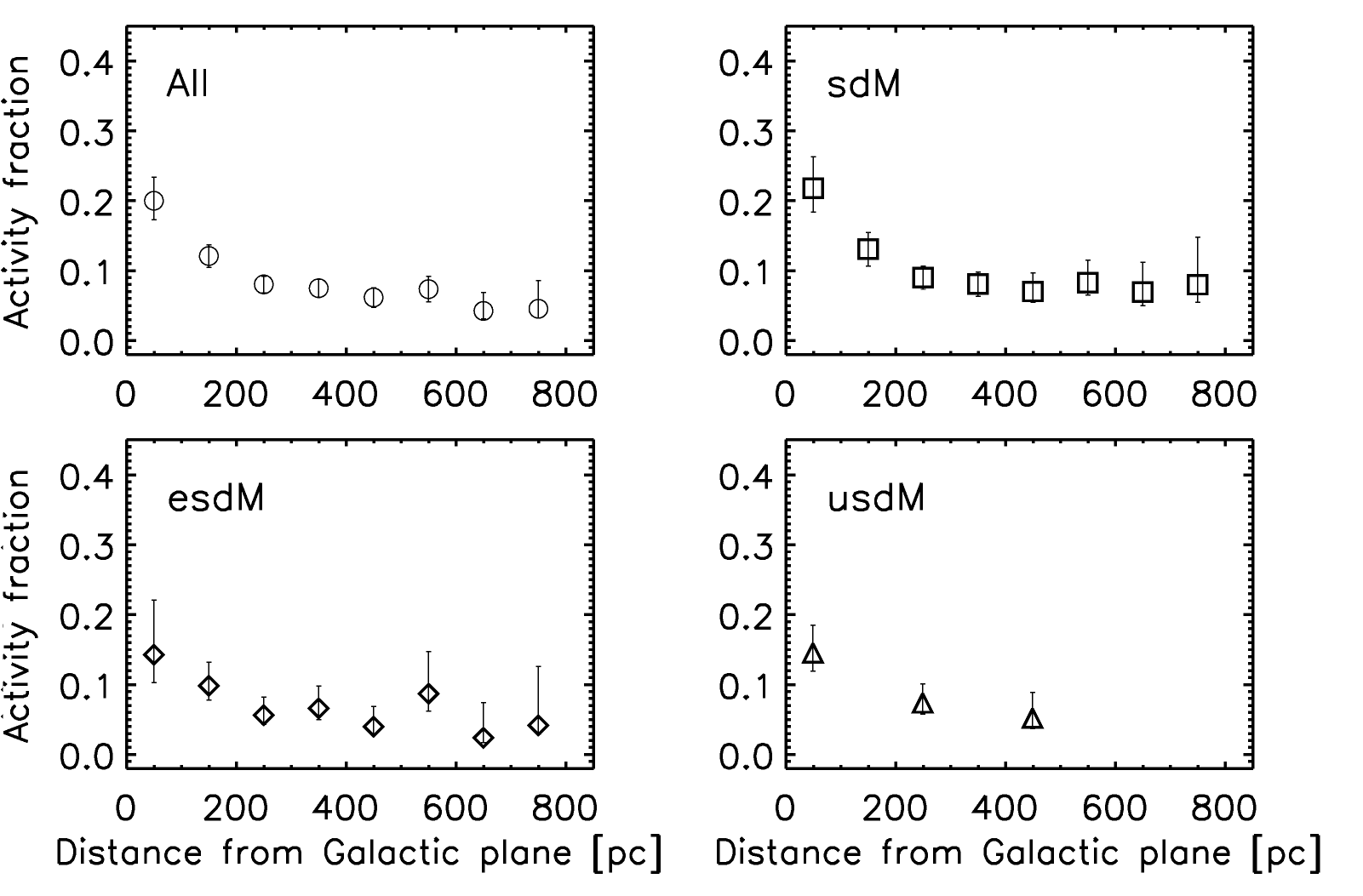}
\caption{Mean activity fractions (number of active stars divided by total number of stars in the bin) in bins of distance from the Galactic plane for all subdwarfs in the sample (upper left) and the three metallicity classes separately (sdMs -- upper right, edMs -- lower left, and usdMs -- lower right). { All definitely active and inactive stars (activity=1 and 0 respectively) from the sample are included.} All spectral classes of the corresponding metallicity class are included in the plots to increase the counting statistics. The error bars derived from binomial statistics are shown as well. There is a strong decrease of the activity fraction with distance from the Galactic plane for sdMs, and a weaker fall off for esdMs and usdMs. \label{actd}}
\end{figure}

\clearpage

\begin{figure}
\epsscale{.90}
\plotone{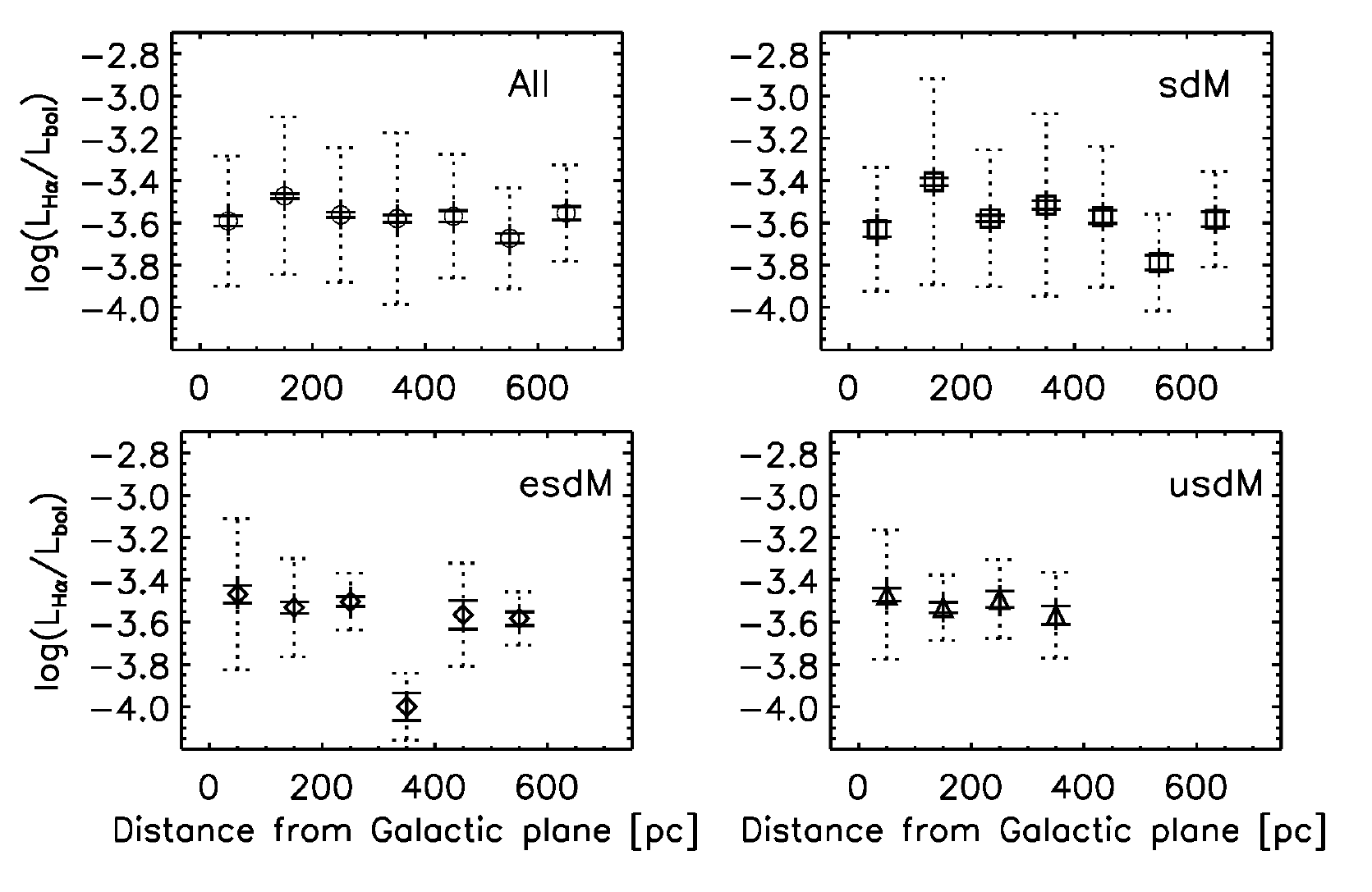}
\caption{Mean values of the luminosity in H${\alpha}$ normalized by the bolometric luminosity ($L_{\mathrm{H\alpha}}/L_{\mathrm{bol}}$) of all subdwarfs in bins of distance from the galactic plane (upper left), for sdMs -- upper right, esdMs -- lower left, usdMs -- lower right. { All active stars from the sample are included.} All spectral classes of the corresponding metallicity class are included in the plots to increase the counting statistics. The error in the mean is plotted as solid error bar, and the standard deviation is given as dotted error bar. { There is no clear trend with distance from the Galactic plane.}\label{lha}}
\end{figure}

\clearpage

\begin{figure}
\epsscale{.90}
\plotone{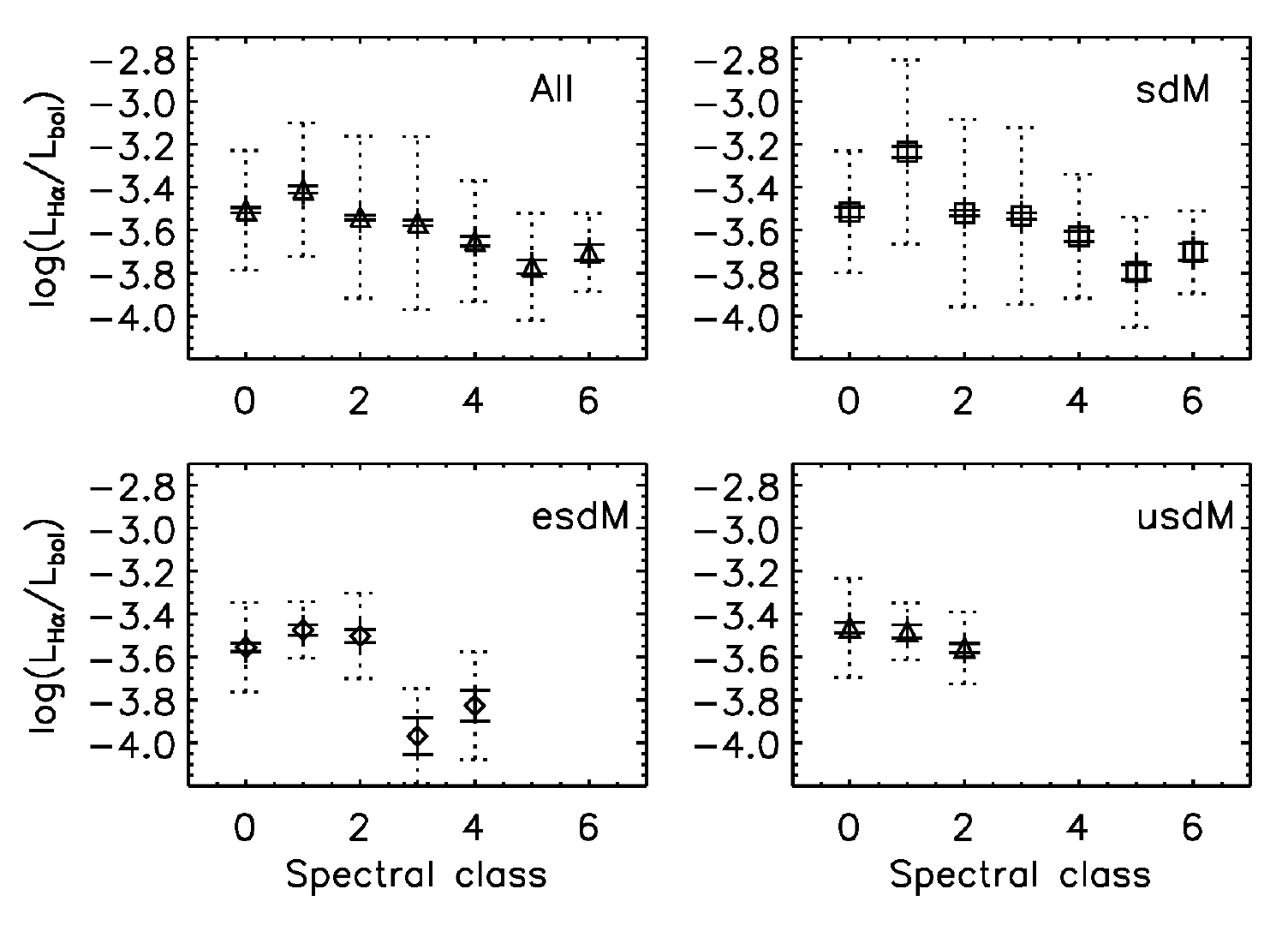}
\caption{Mean values of the luminosity in H${\alpha}$ normalized by the bolometric luminosity ($L_{\mathrm{H\alpha}}/L_{\mathrm{bol}}$) of all subdwarfs in bins of spectral class (upper left), for sdMs -- upper right, esdMs -- lower left, usdMs -- lower right. { Active stars stars from the sample are included.} The error in the mean is plotted as solid error bar, and the standard deviation is given as dotted error bar. { There is a decrease} of the normalized H${\alpha}$ luminosity with spectral class only for the sdMs and no clear trend for the other metallicity classes. \label{lha-sp}}
\end{figure}


\begin{thebibliography}{}
\bibitem[Bessell(1982)]{bessell82} Bessell, M S., 1982, PASAu, 4, 417
\bibitem[Binney \& Merrifield(1998)]{binney98} Binney, J., \& Marrifield, M., 1998, Galactic Astronomy
\bibitem[Bochanski et al.(2007)]{bochanski07} Bochanski, J., West, A., Hawley, S. \& Covey, K., 2007, \aj, 133, 531
\bibitem[Bochanski et al.(2010)]{bochanski10} Bochanski, J. et al., 2010, \aj, 139, 2679
\bibitem[Bochanski et al.(2013)]{bochanski12} Bochanski, J., Savcheva, A., West, A. \& Hawley, S., 2013, \aj, 145, 40
\bibitem[Boeshaar(1976)]{boeshaar76} Boeshaar, P. C., 1976, Ph.D. Thesis, The Ohio State University
\bibitem[Burgasser et al.(2007)]{burgasser07} Burgasser, A., Cruz, K. \& Kirkpatrick, D., 2007, \apj, 657, 494
\bibitem[Burgasser \& Kirkpatrick(2006)]{burgasser06} Burgasser, A. J. \& Kirkpatrick, J. D., 2006, \apj, 645, 1485
\bibitem[Carney \& Latham(1987)]{carney87} Carney, B. \&  Latham, D., 1987, IAUS, 117, 39
\bibitem[Carney et al.(1994)]{carney94} Carney, B. et al., 1994, \aj, 107, 224
\bibitem[Casertano et al.(1990)]{casertano90} Casertano, S., Ratnatunga, K., Bahcall, J., 1990, \apj, 357, 435
\bibitem[Covey et al.(2008)]{covey08} Covey, K., Hawley, S. \&  Bochanski, J. 2008, \aj, 136, 1778
\bibitem[Costa et al.(2005)]{costa05} Costa, E. et al., 2005, \aj, 130, 337
\bibitem[Dawson \& De Robertis(1988)]{dawson88} Dawson, P. C \& De Robertis, M. M., 1988, \aj, 95, 1251
\bibitem[Dhital et al.(2012)]{dhital12} Dhital, S. et al., 2012, \aj, 143, 67
\bibitem[Espinoza Contreras et al.(2013)]{espinosa13} Espinosa Contreras, M. et al., 2013, MnSAI, 84, 963
\bibitem[Folkes et al.(2012)]{folkes12} Folkes, S. L. et al., 2012, MNRAS, 427, 3280
\bibitem[Faherty et al.(2009)]{faherty09} Faherty, J. et al., 2009, \aj, 137, 1F
\bibitem[Fuchs et al.(2009)]{fuchs09} Fuchs, B. et al., 2009, \aj, 137, 4149
\bibitem[Gizis(1997)]{gizis97} Gizis, J., 1997, \aj, 113, 2
\bibitem[Gizis et al.(2002)]{gizis02} Gizis, J., Reid, I. \& Hawley, S., 2002, \aj, 123, 3356
\bibitem[Hambly et al.(2001)]{hambly01} Hambly, N., Irwin, M. \& MacGillivray, H. 2001, MNRAS, 326, 1295
\bibitem[Hartwick et al.(1984)]{hartwick84} Hartwick, P. D, Cowley, A. P. \& Mould, J. R., 1984, \apj, 286, 269
\bibitem[Hawley et al.(1986)]{hawley86} Hawley, S.,Jefferys, W., Barnes, T. III, Lai, W., 1986, \apj, 302, 626 
\bibitem[Hawley et al.(1996)]{hawley96} Hawley, S., Gizis, J. \& Reid, I., 1996, \aj, 112,2799 
\bibitem[Ivezic et al.(2008)]{ivezic08}	Ivezi\'c, Z. et al., 2008, \apj, 684, 287
\bibitem[Jao et al.(2005)]{jao05} Jao, W.-C. et al., 2005, \aj, 129, 1954
\bibitem[Jao et al.(2008)]{jao08} Jao, W.-C. et al., 2008, \aj, 136, 840
\bibitem[Jao et al.(2011)]{jao11} Jao, W.-C. et al., 2011, \aj, 141, 117
\bibitem[Kerber et al.(2001)]{kerber01} Kerber, L., Javiel, S. \& Santiago, B., 2001, A\& A, 365, 424
\bibitem[Kirkpatrick(1992)]{kirkpatrick92} Kirkpatrick, J. D., 1992, Ph.D. Thesis, Univ. Arizona, Tucson
\bibitem[Kirkpatrick et al.(2010)]{kirkpatrick10} Kirkpatrick, J. D. et al., 2010, ApJS, 190, 100\\
\bibitem[Koen(1992)]{koen92} Koen, C., 1992, MNRAS, 265, 65
\bibitem[Kowalski et al.(2009)]{kowalski09} Kowalski, A., Hawley, S. \& Hilton, E., 2009, \aj, 138, 633
\bibitem[Laughlin et al.(1997)]{laughlin97} Laughlin, G., Bodenheimer, P. \& Adams, F. 1997, \apj, 482, 420
\bibitem[L\'epine et al.(2003)]{lepine03} L\'epine, S. et al. 2003, \aj, 125, 1598
\bibitem[L\'epine(2005)]{lepine05a} L\'epine, 2005, \aj, 130, 1247
\bibitem[L\'epine \& Shara(2005)]{lepine05} L\'epine, S. \& Shara, M., 2005, \aj, 129, 1483
\bibitem[L\'epine et al.(2007)]{lepine07} L\'epine, S., Rich, M. \& Shara, M., 2007, \apj, 669, 1235
\bibitem[L\'epine \& Scholz(2008)]{lepine08} L\'epine, S. \& Scholz, R.-D., 2008, \apj, 681, 31
\bibitem[L\'epine et al.(2012)]{lepine12} Lepine, S. et al., 2012, ArXiv e-print, 1206.5991
\bibitem[Lodieu et al.(2012)]{lodieu12} Lodieu, N. et al., 2012, A\& A, 542, 105
\bibitem[Luyten(1922)]{luyten22}Luyten, W. J., 1922, Lick Observatory Bulletin, 10, 135
\bibitem[Majaess(2009)]{majaess09}Majaess, D. J., Turner, D. G. \& Lane, D. J, 2009, MNRAS, 398, 263
\bibitem[Mann et al.(2013)]{mann13} Mann, A. et al., 2013, \aj, 145, 52
\bibitem[Monet et al.(1992)]{monet92} Monet, D. G. et al., 1992, \aj, 103, 638
\bibitem[Monteiro et al.(2006)]{monteiro06} Monteiro, H. et al., 2006, \apj, 638, 446
\bibitem[Mould(1976)]{mould76} Mould, J. R., 1976, \apj, 210, 402
\bibitem[Mould \& McElroy(1978)]{mould78} Mould, J. R. \& McElroy, D. B., 1978, \apj, 220, 935
\bibitem[Munn et al.(2004)]{munn04} Munn, J. A. et al., 2004, \aj, 127, 3034
\bibitem[Munn et al.(2008)]{munn08} Munn, J. A. et al., 2008, \aj, 136, 895
\bibitem[Morgan et al.(2012)]{morgan12} Morgan, D. et al., 2012, \aj, 144, 93
\bibitem[Murray(1983)]{murray83} Murray, C. A. (ed.), 1983, Vectorial Astrometry
\bibitem[Reid et al.(1995)]{reid95} Reid, I., Hawley, S. \& Gizis, J., 1995, \aj, 110, 1838
\bibitem[Reid \& Hawley(2005)]{reid05} Reid, I., Hawley, S., 2005, New light on dark stars, Praxis Publishing
\bibitem[Reid \& Gizis(2005)]{reid05a} Reid, I. \& Gizis, J., 2005, PASP, 117, 676
\bibitem[Reid et al.(2005)]{reid05b} Reid, I., Hawley, S. \& Gizis, J., 2005, \aj, 110, 1838
\bibitem[Reid et al.(2008)]{reid08} Reid, I. et al., 2008, \aj, 136, 2222
\bibitem[Ryan \& Norris(1991a)]{ryan91a} Ryan, S. G. \& Norris, J. E., 1991a, \aj, 101, 1865
\bibitem[Ryan \& Norris(1991b)]{ryan91b} Ryan, S. G. \& Norris, J. E., 1991b, \aj, 101, 1835
\bibitem[Ryan et al.(1991)]{ryanetal91} Ryan, S. G., Norris, J. E. \& Bessell, M. S., 1991, \aj, 102, 303
\bibitem[Schmidt et al.(2010)]{schmidt10} Schmidt, S., West, A., Hawley, S. \& Pineda, J. 2010,\aj, 139, 1808
\bibitem[Schlegel et al.(1998)]{schlegel98} Schlegel, Finkbeiner, \& Davis, 1998, \apj, 500, 525
\bibitem[Sch\"onrich et al.(2010)]{sch10} Sch\"onrich, R., Binney, J. \& Dehnen, W., 2010, MNRAS, 403, 1829
\bibitem[Skrutskie et al.(2006)]{skrutskie06} Skrutskie, M. F. et al., 2006, \aj, 131, 1163\\
\bibitem[Subasavage et al.(2005a)]{subasavage05a} Subasavage, J. P. et al., 2005a, \aj, 129, 413
\bibitem[Subasavage et al.(2005b)]{subasavage05b} Subasavage, J. P. et al., 2005b, \aj, 130, 1658
\bibitem[Tonry \& Davis(1979)]{td79} Tonry J. \& Davis, M., 1979, \aj, 84, 1511
\bibitem[Walkowicz et al.(2004)]{walkowicz04} Walkowicz, L., Hawley, S. \& West, A., 2004, PASP, 116, 1105
\bibitem[West et al.(2004)]{west04} West, A. A. et al., 2004, \aj, 128, 426
\bibitem[West et al.(2006)]{west06} West, A. A. et al., 2006, \aj, 132, 2507
\bibitem[West et al.(2008)]{west08} West, A., Hawley, S. \& Bochanski, J. 2008, \aj, 135, 785
\bibitem[West \& Hawley(2008)]{westH08} West, A. \& Hawley, S., 2008, PASP, 120, 1161
\bibitem[West et al.(2011)]{west11} West, A. A. et al., 2011, \aj, 141, 97
\bibitem[Wing et al.(1976)]{wing76} Wing, R. F., Dean, C. A. \& MacConnell, D. J., 1976, \apj, 205, 186
\bibitem[York et al.(2000)]{york00} York, D. et al., 2000, \aj, 120, 1579
\bibitem[Zhang et al.(2009)]{zhang09} Zhang, Z. H. et al., 2009, A\& A, 497, 619
\end{thebibliography}
\end{document}